\documentclass[a4paper,10pt]{article}
\usepackage[english]{babel}
\usepackage[T1]{fontenc}
\usepackage[utf8]{inputenc}
\usepackage{amssymb}
\usepackage{physics}
\usepackage{amsmath}
\usepackage{graphicx}
\usepackage[unicode]{hyperref}
\usepackage{tikz}
\usepackage{color}
\usepackage{genyoungtabtikz}
\usepackage[title]{appendix}
\usepackage{indentfirst}
\usepackage{bm}
\usepackage{xhfill}
\usepackage{bbm}
\usepackage{multirow}
\usepackage{array}
\usepackage[centertableaux]{ytableau}
\usepackage{multicol}
\usepackage[shortlabels]{enumitem}
\usepackage{forest}
\usepackage{float}

\hypersetup{
    colorlinks=true,
    linkcolor=blue,
    filecolor=magenta,      
    urlcolor=cyan,
}
 
\usepackage{amsthm}

\theoremstyle{definition}

\usepackage[nottoc]{tocbibind}

\usepackage{authblk}

\title{\textbf{From Hurwitz numbers to Feynman diagrams: counting rooted trees in log gravity}}

\author[1,2]{\textbf{Yannick Mvondo-She}}
\affil[1]{School of Physics and Mandelstam Institute for Theoretical Physics,
University of the Witwatersrand, Johannesburg, Wits, 2050, South Africa}
\affil[2]{DSI-NRF Centre of Excellence in Mathematical and Statistical Sciences (CoE-MaSS), South Africa}
\affil[ ]{\texttt{vondosh7@gmail.com}}

\date{}

\begin{document}

\maketitle

\begin{abstract}
We show that the partition function of the logarithmic sector of critical topologically massive gravity which represents a series expansion of composition of functions, can be expressed as a sum over rooted trees. Our work brings a connection between integrable hierarchies of mathematical physics, combinatorial Hopf algebras and rooted trees, by explaining how the $\tau$-functions of the (potential) Burgers and KP integrable hierarchies appearing in the partition function of log gravity conceal the Hopf algebra of composition of functions, known as the Fa\`a di Bruno algebra, of the same type as the celebrated Connes-Kreimer Hopf algebra of rooted trees and Feynman diagrams. In particular, the Hurwitz numbers appearing in the partition function arise as coefficients of isomorphism classes of rooted trees. A parallel is drawn between our findings and established results in the statistical physics literature concerning certain systems with quenched disorder on trees, associated to nonlinear partial differential equations admitting traveling wave solutions. This should be of particular interest in view of a further description of the disorder observed in log gravity. 
\end{abstract}

\tableofcontents

\section{Introduction}
The study of gravity in three dimensions holds the prospect of shedding light on many intricate aspects of classical and quantum gravity. An important result in the study of its asymptotics which anticipated the AdS/CFT correspondence \cite{Maldacena:1997re}, was the emergence of a Virasoro algebra suggesting a dual conformal field theory (CFT) in two dimensions at the boundary \cite{Brown:1986nw}. Pure Einstein gravity in three dimensions is however locally trivial at the classical level and does not exhibit propagating degrees of freedom, hence the need to modify it. One way of doing so is by introducing a negative cosmological constant. Although the resulting theory still has no propagating degrees of freedom, it has black hole solutions \cite{Banados:1992wn}. Another possible modification is to add a gravitational Chern-Simons term. In that case the theory is called topologically massive gravity (TMG), and contains a propagating degree of freedom, the massive graviton \cite{Deser:1982vy,Deser:1981wh}. When both cosmological and Chern-Simons terms are included in a theory, it yields cosmological topologically massive gravity (CTMG). Such a theory features both gravitons and black holes.

A proposal to find a CFT dual to Einstein gravity \cite{Witten:2007kt} was made in 2007, and was followed by the calculation of the graviton 1-loop partition function \cite{Maloney:2007ud}. However, discrepancies were found in the results, in particular with the failing in factorization of the left- and right-contributions, therefore clashing with the proposal of \cite{Witten:2007kt}. Soon after, a non-trivial slightly modified version of the aforementioned proposal was formulated \cite{Li:2008dq}, in which Einstein gravity was replaced by chiral gravity, which can be viewed as a special case of topologically massive gravity at a specific tuning of the couplings, and is asymptotically defined with AdS$_3$ boundary conditions, in the spirit of Fefferman-Graham-Brown-Henneaux \cite{Brown:1986nw,graham1985charles,Maloney:2009ck}. A particular feature of the theory was that one of the two central charges vanishes, whilst the other one can have a non-zero value. This gave an indication that the partition function could factorize.

Shortly after the proposal of \cite{Li:2008dq}, Grumiller \textit{et al.} noticed that relaxing the Brown-Henneaux boundary conditions allowed for the presence of a massive mode that forms a Jordan cell with the massless graviton, leading to a degeneracy at the critical point \cite{Grumiller:2008qz}. In addition, it was observed that the presence of the massive mode spoils the chirality of the theory, as well as its
unitarity. The non-unitarity arising from the appearance of the Jordan cell is a salient feature in logarithmic conformal field theories (LCFTs), which led to the conjecture that the dual CFT of critical cosmological topologically massive gravity (cTMG) should be an LCFT with central charge $c=0$ \cite{Cardy:1999zp,Gurarie:1999yx,Gurarie:2004ce}. The massive mode was called the logarithmic partner of the graviton, and the theory gravity with logarithmically relaxed boundary conditions in \cite{Grumiller:2008qz} is also referred to as \textit{log gravity}. 

The proposed non-unitary holography received further support through the calculation of correlation functions in TMG \cite{Skenderis:2009nt,Grumiller:2009mw}, which confirmed the existence of logarithmic correlators of the type that arises in LCFT. Subsequently, the 1-loop graviton partition function of cTMG  on the thermal AdS$_3$ background was calculated in \cite{Gaberdiel:2010xv}, resulting in the following expression

\begin{eqnarray}
\label{z tmg}
{Z_{\rm{cTMG}}} (q, \bar{q})= \prod_{n=2}^{\infty} \frac{1}{|1-q^n|^2} \prod_{m=2}^{\infty} \prod_{\bar{m}=0}^{\infty} \frac{1}{1-q^m \bar{q}^{\bar{m}}}, \hspace{1cm} \text{with} \hspace{0.25cm} q=e^{2i \pi \tau}, \bar{q}=e^{-2i \pi \bar{\tau}},
\end{eqnarray}
where the first product can be identified as the three-dimensional gravity partition function $Z_{0,1}$ in \cite{Maloney:2007ud}. In search for a better insight into the 1-loop partition function as proposed in \cite{Grumiller:2013at}, it was shown in \cite{Mvondo-She:2018htn} that the partition function of critical cosmological TMG can be expressed in terms of Bell polynomials. Writing Eq. (\ref{z tmg}) as 

\begin{eqnarray}
\label{z grav z log}
{Z_{\rm{cTMG}}} (q, \bar{q})=  {Z_{\rm{gravity}}} (q, \bar{q}) \cdot {Z_{\rm{log}}} (q, \bar{q}),
\end{eqnarray}
where

\begin{eqnarray}
{Z_{\rm{gravity}}} (q, \bar{q})= \prod_{n=2}^{\infty} \frac{1}{|1-q^n|^2}, \hspace{0.5cm} \mbox{and}
\hspace{0.5cm} {Z_{\rm{log}}} (q, \bar{q}) = \prod_{m=2}^{\infty} \prod_{\bar{m}=0}^{\infty} \frac{1}{1-q^m \bar{q}^{\bar{m}}},
\end{eqnarray}

\noindent it was shown that the log partition function ${Z_{\rm{log}}} (q, \bar{q})$ is the generating function of Bell polynomials $Y_n \left( g_1, \ldots, g_n \right)$

\begin{eqnarray}
\label{z log bell}
{Z_{\rm{log}}} (q, \bar{q}) = \sum_{n=0}^{\infty} \frac{Y_n}{n!} \left( q^{2} \right)^n, 
\end{eqnarray}

\noindent where 

\begin{eqnarray}
g_n = (n-1)! \sum_{m \geq 0, \bar{m} \geq 0}  q^{nm} \bar{q}^{n \bar{m}} = (n-1)! \frac{1}{|1-q^n|^2}.
\end{eqnarray}

\noindent Under the rescaling of variables

\begin{eqnarray}
g_n (q,\bar{q})= (n-1)! \mathcal{G}_{n} (q,\bar{q}),
\end{eqnarray}

\noindent where 

\begin{eqnarray}
\label{G_n}
\mathcal{G}_n \left( q,\bar{q} \right) = \frac{1}{|1-q^n|^2}, 
\end{eqnarray} 

\noindent the log partition function function was also shown to be expressed in the form of the bosonic plethystic exponential $PE^{\mathcal{B}}$ as 

\begin{eqnarray}
\label{PE}
Z_{\rm{log}} \left(  q,\bar{q} \right) = PE^{\mathcal{B}} \left[ \mathcal{G}_{1} (q,\bar{q}) \right] =  \exp \left( \sum^{\infty}_{n=1} \frac{\left( q^2 \right)^n}{n}  \mathcal{G}_n \left( q,\bar{q} \right) \right).
\end{eqnarray}

Recently, further results \cite{Mvondo-She:2021joh} were obtained in terms of properties of the log partition function. On one hand, a structure related to soliton solutions of the KP (Kadomtsev-Petviashvili) hierarchy was found in the log partition function. The KP hierarchy is a completely integrable hierarchy, i.e a family of partial differential equations which are simultaneously solved, that generalizes the KdV (Korteveg-de-Vries) hierarchy. Using the Sato universal Grassmannian which is the moduli space of the formal power series solutions of the KP hierarchy \cite{sato1981soliton}, the log partition function was identified as a $\tau$-function of the KP hierarchy. Over the last three decades, a strong interest in the connection between integrable hierarchies and moduli spaces of curves has grown, and $\tau$-functions have played an important role in this relationship, starting with Witten’s Conjecture \cite{Witten:1990hr} proved by Kontsevich \cite{Kontsevich:1992ti}, which showed that the $\tau$-function of a special solution of the KdV hierarchy generates intersection indices of certain cohomology classes on moduli spaces of curves. On the other hand, the log partition function was found to satisfy the potential form of the Burgers hierarchy \cite{Adler:2015vya,Adler:2015wft}, which is a family formed by an infinite tower of nonlinear differential equations whose first member is the Burgers equation, originally used to describe the mathematical modelling of turbulence \cite{Burgers1948AMM}. 

It was also shown that by generating Hurwitz numbers, the log partition function is a generating function of the combinatorics of branched covering maps. Hurwitz theory studies maps of Riemann surfaces by enumerating analytic functions between the Riemann surfaces \cite{cavalieri2016riemann}. The study of analytic functions on Riemann surfaces translates into the study of the geometry of oriented topological surfaces, with the analytic functions being identified as ramified coverings. The number of such functions fixed by the appropriate set of discrete invariants is counted by Hurwitz numbers. The latter thus enumerate ramified coverings of Riemann surfaces.

Ramified coverings naturally induce monodromy representations, i.e homomorphisms from the fundamental group of the punctured target surface to a symmetric group. The ramifications at the preimages of a point in the base surface is captured by the cycle type of permutations, making it possible to obtain closed formulae for Hurwitz number in terms of characters of the symmetric group. As a consequence, the work in \cite{Mvondo-She:2021joh} showed an equivalence between the problem of counting the number of ways a genus zero Riemann surface can be covered $n$-times with branch points allowed, and computing the partition function of a gauge theory defined on a Riemann surface, with the symmetric group $S_n$ as the gauge group, where the moduli space $\left( \mathbb{C}^2 \right)^n /S_n$ is a Hurwitz space of branched coverings. The log partition function can therefore be described as a Hurwitz $\tau$-function, on one hand computing Hurwitz numbers and on the other hand being a $\tau$-function of the KP hierarchy. Hurwitz $\tau$-functions are a relatively new important subject in theoretical physics, having also appeared in the literature as partition functions for HOMFLY polynomials of some knot in the theory of Ooguri-Vafa (OV) \cite{Mironov:2013gza,Mironov:2013oma,Sleptsov:2014aba}. In particular, for any torus knot, the OV partition function is a $\tau$-function of the KP hierarchy \cite{Mironov:2011ym}. This shows a deep connection between Hurwitz $\tau$-functions and 3d Chern-Simons theory, which also appears in this Chern-Simons formulation of a non-unitary three-dimensional gravity.

The purpose of this paper is in the first place to show that the log partition function can be expressed as a sum of contributions indexed by isomorphism classes of rooted trees. This is done by describing how the Fa\`a di Bruno Hopf algebra that is of the same type as a Hopf algebra of Feynman graphs, can be constructed from the Bell polynomials that make up the log partition function. In so doing, our work connects the results obtained in \cite{Mvondo-She:2021joh} to the Connes-Kreimer Hopf algebra of rooted trees and Feynman diagrams, providing further evidence of the relation between the Hopf algebraic structures introduced by Connes and Kreimer and the integrable structure of $\tau$-functions \cite{Gerasimov:2000pr}. At the same time, it is shown that Hurwitz numbers appear in the partition function as coefficients in the sum over rooted trees. The second purpose of this paper is to bring a connection between the aforementioned results and statistical physics models with quenched disorder on trees, as a way for further interpretation of the disorder introduced in AdS$_3$ leading to the theory of log gravity. This paper is organized as described below.

 In section 2, we give a description of the Hopf algebraic structure that underlies the $tau$-functions of the (potential) Burgers and the KP integrable hierarchies. We start by giving a brief definition of Hopf algebras. We then introduce the Hopf algebra of composition of functions, known as the Fa\`a di Bruno algebra, and show how the Bell polynomials in the log partition function give rise to such Hopf algebraic structure.

In section 3, we spend time on the reformulation of the logarithmic partition function as a sum over rooted trees. We start by discussing the relationship between rooted trees and linear differential operators from a Hopf algebra perspective. We then establish the correspondence between the differential terms arising in higher order derivatives of a composition of two functions and rooted trees. After that, in order to make more contact with physics, we make more explicit the expansion of the log partition function using the language of renormalization in quantum field theory. From a physical perspective, phenomena at very small length scales (i.e at very large energy scales) are described by quantum field theory (QFT). Despite the many successes of QFT, its mathematical construction is plagued by the difficulty of computing quantities in integral form without incurring infinities. A palliative treatment called \textit{renormalization} has been applied on the perturbative expansions of divergent iterated Feynman integrals to render renormalized values. While looking for the mathematical structure behind the renormalization method of quantum field theory, a remarkable mathematical interpretation of the perturbative renormalization was discovered by Kreimer \cite{Kreimer:1997dp}, in arranging Feynman diagrams of the renormalizable QFT into a Hopf algebra. Subsequent works by Kreimer and Connes led to a reformulation of many quantum field constructions such as renormalization in a Hopf algebraic language, placing Hopf algebras at the heart of the noncommutative approach to geometry and physics \cite{Connes:1998qv,Connes:1999yr,Connes:2000fe}. In their approach to the renormalization of perturbative QFT, a major role is played by an algebraic structure defined over a set of Feynman diagrams called the Hopf algebra of rooted trees. The latter has a subalgebra whose generators will be used to derive an expression of the log partition function in terms of rooted trees.
Because we are primarily interested in the counting of trees generated by random permutation, we derive an explicit expression of the log partition function in terms of rooted trees indexed by Hurwitz numbers, and show the cycle decomposition of permutations on the vertices of the rooted trees.

In section 4, we draw a parallel between our findings and results well established in the statistical physics literature, concerning disordered systems on trees. In the particular case of the directed polymer on a tree with disorder, it has been shown that the study of such a disordered system reduces to that of nonlinear partial differential equations admitting travelling wave solutions \cite{derrida1988polymers}. The analogy is of interest in view of a further elucidation of phenomena associated to the disorder introduced in AdS$_3$ as observed in \cite{Grumiller:2008qz}.

In section 5, we give a summary and outlook.

\section{Hopf algebraic structure in the $\tau$-functions of the (potential) Burgers and the KP integrable hierarchies}
Hopf algebras have been used for a long time in an implicit way in statistical physics and quantum field theory. Indeed, in \cite{Stora:1973dka} and \cite{Borchers:1975cu} for instance, the authors made use of a product called the convolution product in the Hopf language. Since the work of Joni and Rota \cite{joni1979coalgebras}, Hopf algebras have also become a sophisticated tool to formalize combinatorics. Indeed, on one hand, the product and the coproduct can capture the actions of composing and decomposing combinatorial objects respectively, and on the other hand, combinatorial objects such as permutations, trees, graphs, posets, or tableaux have natural gradings which allow them to be endowed with a Hopf algebraic structure, such that many interesting invariants can be expressed as Hopf morphisms.

We start this section by giving a background of algebraic concepts that are important in the paper. We thereafter introduce the Fa\`a di Bruno Hopf algebra and give its description as a Hopf algebra on a polynomial ring whose coproduct is expressed in terms of Bell polynomials.

\subsection{Hopf algebra résumé}
Following \cite{zeidler2008quantum}, we briefly review the definition of a Hopf algebra. An associative complex algebra $\mathcal{A}$ with unit element \textbf{1} is a $\mathbb{C}$-vector space given for all $a,b \in \mathcal{A}$ and all complex numbers $z$ by the following maps

\begin{subequations}
\begin{align}
& \mbox{Multiplication map: } f (a \otimes b) = ab, \\
&\mbox{Unit map: } g(z)=z\bf{1}, \\
&\mbox{Identity map: } \mbox{id} (a) =a.
\end{align}
\end{subequations}

\noindent Using linear extensions, the $\mathbb{C}$-linear maps of the algebra $\mathcal{A}$ can be written

\begin{eqnarray}
f: \mathcal{A} \otimes \mathcal{A} \mapsto \mathcal{A}, \hspace{1cm}
g: \mathbbm{C} \mapsto \mathcal{A}, \hspace{1cm}
\mbox{id:} \mathcal{A} \rightarrow \mathcal{A},
\end{eqnarray}

\noindent and provided with associativity and unity properties 

\begin{subequations}
\begin{align}
 & \mbox{Associativity: } f(f \otimes \mbox{id}) = f(\mbox{id} \otimes f), \label{associativity} \\
 & \mbox{Unity: } f(g \otimes \mbox{id})= f(\mbox{id} \otimes g)= \mbox{id}. \label{unity}
\end{align}
\end{subequations}

\noindent The associativity relation (\ref{associativity}) following from the associative law $a(bc)=(ab)c$ for all $a,b,c \in \mathcal{A}$, one can write

\begin{eqnarray}
f(\mbox{id} \otimes f) (a \otimes b \otimes c) = f ( a \otimes f(b \otimes c)) = f(a \otimes bc) = a(bc).
\end{eqnarray}

\noindent and similarly $f(f \otimes \mbox{id}) (a \otimes b \otimes c)= (ab)c$. The unity relation (\ref{unity}) follows from 

\begin{eqnarray}
f(g \otimes \mbox{id})(1 \otimes a) = f( g(1) \otimes a) = f ( {\bf{1}} \otimes a) = {\bf{1}} a = a.
\end{eqnarray}

\noindent In the same way, $f( \mbox{id} \otimes g)(a \otimes 1) = a {\bf{1}} = a$, and the relation (\ref{unity}) is then proved by identifying $a \otimes 1$ and $1 \otimes a$ with $a$, which corresponds to the isomorphisms $\mathcal{A} \otimes \mathbbm{C} = \mathcal{A} = \mathbbm{C} \otimes \mathcal{A}$.

\noindent The associative unital complex algebra $\mathcal{A}$ becomes a complex bialgebra if there exist two maps

\begin{subequations}
\begin{align}
& \mbox{Comultiplication: } \Delta: \mathcal{A} \rightarrow \mathcal{A} \otimes \mathcal{A}, \\
& \mbox{Counit: } \varepsilon: \mathcal{A} \rightarrow \mathbbm{C},
\end{align}
\end{subequations}

\noindent satisfying the compatibility condition ensured by requiring $\Delta$ and $\varepsilon$ to be unital algebra homomorphisms, and satisfying the coassociativity and the counity properties 

\begin{subequations}
\begin{align}
& \mbox{Coassociativity: } (\mbox{id} \otimes \Delta) \Delta = (\Delta \otimes \mbox{id}) \Delta, \\
& \mbox{Counity: } (\mbox{id} \otimes \varepsilon) \Delta = (\varepsilon \otimes \mbox{id})\Delta = \mbox{id}.
\end{align}
\end{subequations}

\noindent The complex bialgebra is called a Hopf algebra  if there exists a linear map $S: \mathcal{A} \rightarrow \mathcal{A}$ such that 

\begin{eqnarray}
f(S \otimes \mbox{id}) \Delta = f( \mbox{id} \otimes S) \Delta = g \varepsilon.
\end{eqnarray}

\noindent For all $a \in \mathcal{A}$, the coproduct $\Delta a$ is contained in the tensor product $\mathcal{A} \otimes \mathcal{A}$. Therefore, there are elements $a_1, \ldots, a_n, b_1, \ldots b_n$ such that 

\begin{eqnarray}
\label{coproduct}
\Delta a = \sum_{k=1}^n a_k \otimes b_k.
\end{eqnarray}

\noindent Using the so-called Sweedler notation, Eq. (\ref{coproduct}) also takes the form 

\begin{eqnarray}
\Delta a = a_{(1)} \otimes a_{(2)}.
\end{eqnarray}

\noindent Furthermore, for all $a,b \in \mathcal{A}$, one can write

\begin{eqnarray}
\Delta a \Delta b = \Delta (ab), \hspace{1cm} \Delta 1 = 1 \otimes 1.
\end{eqnarray}

Let $\mathbb{C}$[[\textit{t}]] be the ring of formal series on $\mathbb{C}$, and $\mathbb{C}$[$t^{-1},t$] the field of Laurent polynomials on $\mathbb{C}$. A graded Hopf algebra is the direct sum of vector spaces

\begin{eqnarray}
\mathcal{A} = \bigoplus_{n \geq 0} \mathcal{A}_n
\end{eqnarray}
endowed with a product $f: \mathcal{A} \otimes \mathcal{A} \mapsto \mathcal{A}$, a coproduct $\Delta: \mathcal{A} \rightarrow \mathcal{A} \otimes \mathcal{A}$, a unit $g: \mathbbm{C} \mapsto \mathcal{A}$, a counit $\varepsilon: \mathcal{A} \rightarrow \mathbbm{C}$ and an antipode $S: \mathcal{A} \rightarrow \mathcal{A}$ fulfilling the usual axioms of a Hopf algebra, and such that 

\begin{subequations}
\begin{align}
    \mathcal{A}_p \mathcal{A}_q &\subset \mathcal{A}_{p+q} \\
    \Delta \left( \mathcal{A}_n \right) &\subset \bigoplus_{p+q=n} \mathcal{A}_p \otimes \mathcal{A}_q, \hspace{0.25cm} n \geq 0, \hspace{0.25cm} 0\leq p,q \leq n\\
    S \left( \mathcal{A}_n \right) &\subset \mathcal{A}_n
\end{align}
\end{subequations}

Finally, a graded Hopf algebra $\mathcal{A}$ is connected if $\mathcal{A}_0$ is one-dimensional, i.e $\mathcal{A} \cong \mathbb{C}$. The Hopf algebras considered in this paper are graded. 


\subsection{Fa\`a di Bruno Hopf algebraic structure in the log partition function}
Fa\`a di Bruno Hopf algebras have been described several times in various branches of mathematics and physics \cite{joni1979coalgebras,Figueroa:2004hb,zeidler2008quantum,frabetti2015five} and can be introduced in many ways. Our starting point is the set $G$ of formal exponential power series of expression 

\begin{eqnarray}
f(x) = \sum_{n=1}^\infty \frac{fn}{n!} x^n,    
\end{eqnarray}

\noindent with $f_0=0$ and $f_1>0$. By defining in the same way the function $h \in G$ as 

\begin{eqnarray}
g(x) = \sum_{n=1}^\infty \frac{g_n}{n!} x^n,    
\end{eqnarray}

\noindent With respect to composition $f \circ g$, the set $G$ becomes a group called the (formal) local diffeomorphism group of the Gaussian plane $\mathbb{C}$ at the origin. Upon "coordinatization" (this term is employed in \cite{zeidler2008quantum}, appropriate complex-valued coordinate functions on $G$ generate a Hopf algebra denoted by $\mathcal{H}(G)$, also called the coordinate algebra of $G$. More specifically, the $n$-th coordinate map $a_n: G \mapsto \mathbb{C}$ which sends the formal power series $f$ and $h$ to their $n$-th complex coefficients $f_n$ and $h_n$ respectively, can be defined by setting 

\begin{eqnarray}
 a_n(f)=f_n, \hspace{0.5cm} a_n(h)=h_n, \hspace{1cm} n=1,2, \ldots   
\end{eqnarray}

\noindent Then, for all indices $n,m = 1,2$, all formal power series $f,h \in G$, and all complex numbers, the Fa\`a di Bruno Hopf Algebra for the formal
diffeomorphism group $G$ of the complex plane is the algebra endowed with the following set of operations

\begin{subequations}
\begin{align}
&\text{Linear combination:}~~ \left( \alpha a_n + \beta a_m \right) \left( f \right) := \alpha a_n \left( f \right) + \beta a_m \left( f \right), \\ 
&\text{Product:}~~ \left( a_n a_m \right) := a_n \left( f \right) a_m \left( f \right)\\  
&\text{Coproduct:}~~ \left( \Delta a_n  \right) \left( g,f \right) := a_n \left(  f \circ g \right),\\
&\text{Counit:}~~ \varepsilon \left( a_n \right) = a_n \left( \bf{1} \right), \\
&\text{Coinverse:}~~ \left( Sa_n  \right) \left(  f \right) := \left( f^{-1}  \right),
\end{align}    
\end{subequations}

\noindent where, given the unit element $\bf{1}$ of $G$ corresponding to the term $x$ in the power series, the counit is more explicitly expressed as

\begin{equation}
\varepsilon \left( g_n \right) = 
\begin{cases}
1 & \text{if $n=1$}, \\ 
0 & \text{otherwise},
\end{cases}
\end{equation}

\noindent and $f^{-1}$ denotes the reciprocal series of $f$. Both coinverse and coproduct can be derived with the help of the partial Bell polynomials $B_{n,k}$, which are the constituents of the complete Bell polynomials generated by the log partition function. Indeed, as a recall, the Bell polynomial $Y_n$ in variables $g_1, g_2, \ldots, g_n$, is the sum over the $n$-tuples of nonnegative integers $j = \left( j_1, j_2, \ldots, j_n \right)$ such that $j_1 + 2 j_2 + \cdots + n j_n = n$, i.e

\begin{eqnarray}
Y_n  \left( g_1, g_2, \ldots, g_n \right)= \sum_{j_1 + 2 j_2 + \cdots + n j_n = n} \frac{n!}{j_1! j_2! \cdots j_n!} \left( \frac{g_1}{1!} \right)^{j_1} \left( \frac{g_2}{2!} \right)^{j_2} \cdots \left( \frac{g_n}{n!} \right)^{j_n},
\end{eqnarray}

\noindent and the partial Bell polynomials $B_{n,k}$ is the sum over the $n$-tuples of nonnegative integers $j = \left( j_1, j_2, \ldots, j_n \right)$ such that $j_1 + 2 j_2 + \cdots + n j_n = n$ and $j_1 +  j_2 + \cdots +  j_n = k$, i.e

\begin{eqnarray}
B_{n,k} \left( g_1, g_2, \ldots, g_{n+1-k} \right) = \sum_{\substack{j_1 + 2 j_2 + \cdots + n j_n = n \\ j_1 +  j_2 + \cdots +  j_n = k }} \frac{n!}{j_1! j_2! \cdots j_n!} \left( \frac{g_1}{1!} \right)^{j_1} \left( \frac{g_2}{2!} \right)^{j_2} \cdots \left( \frac{g_n}{n!} \right)^{j_n}.
\end{eqnarray}

\noindent The coinverse is expressed in terms of the partial Bell polynomials as follows \cite{Figueroa:2004hb}

\begin{eqnarray}
Sa_n = -a_n + \sum_{j=2}^{n-1} \left( -1 \right)^j   \sum_{1 < k_{j-1} < \cdots < k_1 < n } B_{n,k_1} B_{k_1,k_2} \cdots B_{k_{j-2}, k_{j-1}} a_{k_{j-1}}.
\end{eqnarray}

\noindent with the following special cases

\begin{subequations}
\begin{align}
Sa_2 &= -a_2, \\
Sa_3 &= -a_3 + 3 A^2_2, \\ 
Sa_4 &= -a_4 + 15 a^3_2 + 10 a_2 a_3.
\end{align}    
\end{subequations}

\noindent Of much interest to us, the coproduct formula takes the general form

\begin{eqnarray}
\Delta a_n = \sum_{n=1}^n B_{n,k} \left( a_1,a_2, \ldots, a_{n+1-k} \right) \otimes a_k, \hspace{1cm} n=1,2,\ldots,    
\end{eqnarray}

\noindent which up to order $n=3$ reads

\begin{subequations}
\begin{align}
 &\Delta a_1 = a_1 \otimes a_1, \\
 & \Delta a_2 = a_2 \otimes a_1 + a_1^2 \otimes a_2, \\
 & \Delta a_3 = a_3 \otimes a_1 + 3 a_1 a_2 \otimes a_2 + a_1^3 \otimes a_3.
\end{align}    
\end{subequations}

\noindent The coincidence between the coproduct and the composition $f \circ g$ can be seen from the expansion 

\begin{eqnarray}
 f \left( g \left( x \right) \right) = g_1 f_1 x + \left( g_2 f_1 + g_1^2 f_2 \right) \frac{x^2}{2!} + \left( g_3f_1 + 3g_1g_2f_2 + g^3_1f_3 \right) \frac{x^3}{3!} + \cdots,  
\end{eqnarray}

\noindent where symbolically, the linear part of the coproduct standing on the right side of the $\otimes$ sign is represented by $f_n$, and the polynomial
part of the coproduct on the left of the $\otimes$ sign is represented by $g_n$. As an example 

\begin{eqnarray}
\Delta a_2 \left( g,f \right) = a_2(g) a_1 (f) + a_1(g)^2 a_2 (f) = g_2f_1 + g_1^2 f_2.  
\end{eqnarray}

From the above exposition, the Fa\`a di Bruno Hopf algebraic structure of the $\tau$-function of the potential Burgers hierarchy has been shown. In the case of the $\tau$-function of the KP hierarchy a Fa\`a di Bruno-like Hopf algebraic structure can be derived as a Hopf algebra of polynomials in variables $b_1, b_2, \ldots$, where the coproduct takes the general form

\begin{eqnarray}
\Delta b_n = \sum_{k=1}^n \left( \sum_j   \frac{1}{j_1! j_2! \cdots j_n!} \left( \frac{b_1}{1!} \right)^{j_1} \left( \frac{b_2}{2!} \right)^{j_2} \cdots \left( \frac{b_n}{n!} \right)^{j_n} \right)  \otimes b_k,   
\end{eqnarray}

\noindent where the polynomial part of the coproduct on the left of the $\otimes$ sign consists of components of the 1-part Schur polynomials.

\section{Log partition function as a sum of rooted trees}
In this section, we discuss Hopf algebraic structures on the space of rooted
trees, and show how these structures are related to series expansions of composition of functions via differential operators associated to the trees. As a start, we briefly describe the Hopf algebra of differential operators with constant coefficients, which  shows that the mere fact of calculating the derivative of a product of two functions can intuitively give some insight on Hopf algebras.

\subsection{The Hopf algebra of differential operators}
Fix the dimension $N=1,2, \ldots$, then denote the points of the space $\mathbb{R}^N$ by $x=\left( x_1, \ldots, x_N \right)$ the space of smooth
complex-valued functions $f: \mathbb{R}^n \mapsto \mathcal{C}$. The algebra $\mathcal{A}$ of linear differential operators with respect to $N$ arguments and constant coefficients is generated by the partial derivatives $\partial_i = \partial / \partial x_i$ \cite{brouder2004trees}. The product emerges naturally when $\mathcal{A}$ acts on functions. For smooth functions $f$, the product is the composition of derivatives $\partial_1 \partial_j f = \partial_1 \partial_j f$, for $i,j= 1,\ldots, N$. Furthermore, $\mathcal{A}$ can be endowed with the unit element $\bf{1}$ such that, for any $D \in \mathcal{A}$, $({\bf{1}} D) f= (D {\bf{1}}) f= Df$. Thus $\mathcal{A}$ is an associative, commutative, and unital complex algebra. In order to make it a Hopf algebra, three objects are required: the coproduct, the counit and the antipode. The coproduct appears when the action of $D$ on a product of smooth functions $fg$ is considered. For instance

\begin{subequations}
\label{HA of DO}
\begin{align}
{\bf{1}}(fg)&=fg=({\bf{1}}f)({\bf{1}}g),  \\  
\partial_i(fg)&=(\partial_if)g +f(\partial_ig)=(\partial_if)({\bf{1}}g)+({\bf{1}}f)(\partial_ig). \label{HA of DO subeq2}
\end{align}
\end{subequations}

\noindent By convention, writing the action of an element $D$ of $\mathcal{A}$ on $fg$ as $D(fg)=\sum \left( D_{(1)} f \right) \left( D_{(2)} g \right)$ where $D_{(1)},D_{(2)} \in \mathcal{A}$, the coproduct arises by omitting reference to the functions $f$ and $g$, and defining the linear map $\Delta \mathcal{A} \mapsto \mathcal{A} \otimes \mathcal{A}$ thus obtaining the so-called Sweedler notation $\Delta D = D_{(1)} \otimes D_{(2)}$. In this case, Eqs. (\ref{HA of DO}) give

\begin{subequations}
\label{HA of DO (2)}
\begin{align}
\Delta {\bf{1}}&={\bf{1}} \otimes {\bf{1}},  \\  
\Delta \partial_i&=\partial_i \otimes {\bf{1}} + {\bf{1}} \otimes \partial_i.
\end{align}
\end{subequations}

\noindent The mathematical correspondence of the processes of fusion and splitting of physical states observed in nature is the product and coproduct
of Hopf algebras, respectively. Here, the coproduct can then be seen as a procedure to split a differential operators $D$ into two operators $ D_{(1)}$ and $D_{(2)}$ such that $D_{(1)} D_{(2)} = D$. A pictorial representation of the process is given below

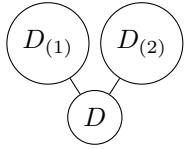
\begin{center}
\begin{forest}
for tree={l sep=0.2em, s sep=0.4em, anchor=center,grow=north,}
[$D$,circle,draw,  [$D_{(2)}$,circle,draw, ] [$D_{(1)}$,circle,draw]] 
\end{forest}   
\end{center}

\noindent An iterated application yields of the product rule in Eq. (\ref{HA of DO subeq2}) yields 

\begin{eqnarray}
\partial_i \partial_j (fg) = (\partial_i \partial_j f)g + \partial_j f \partial_i g + \partial_i f \partial_j g + f \partial_i \partial_j g,
\end{eqnarray}

\noindent from which we get 

\begin{eqnarray}
\Delta (\partial_i \partial_j)(fg)= (\partial_i \partial_j \otimes {\bf{1}})(fg) + (\partial_j \otimes \partial_i)(fg) + (\partial_i \otimes \partial_j)(fg) + ({\bf{1}} \otimes \partial_i \partial_j )(fg).
\end{eqnarray}

\noindent Hence, the coproduct 

\begin{eqnarray}
\Delta (\partial_i \partial_j) = \partial_i \partial_j \otimes {\bf{1}} + \partial_j \otimes \partial_i + \partial_i \otimes \partial_j + {\bf{1}} \otimes \partial_i \partial_j
\end{eqnarray}

\noindent corresponds to the four possible ways of splitting of the product $\Delta (\partial_i \partial_j)$. Using the more compact Sweedler notation, if $\Delta D = D_{(1)} \otimes D_{(2)}$ and $\Delta D' = D'_{(1)} \otimes D'_{(2)}$, then the relation between the product and the coproduct in $\mathcal{A}$ is $\Delta (DD')= (\Delta D) \left(\Delta D' \right) = \sum \sum \left( D_{(1)} D'_{(1)} \right) \otimes \left( D_{(2)}  D'_{(2)} \right)$.

The coassociativity of the coproduct $\Delta$ comes from the associative law $f(gh)=(fg)h$ for functions $f, g, h$, from which follows $D \left(f(gh) \right) = D \left( (fg) h \right)$. In turn, from $D(fgh)= \sum \left( D_{(1)} f \right) \left( D_{(2)} g \right) \left( D_{(3)} h \right)$, one can write 

\begin{eqnarray}
\sum D_{(1)} \otimes D_{(2)} \otimes  D_{(3)} = \left(  \Delta \otimes \rm{id} \right) \Delta D = \left( \rm{id} \otimes \Delta   \right) \Delta D.
\end{eqnarray}

The remaining ingredients for $\mathcal{A}$ to be defined as a Hopf algebra are the counit $\varepsilon$ and the antipode $S$. In the case of the algebra of differential operators the counit can be defined as the linear map $\varepsilon : \mathcal{A} \mapsto \mathbb{C}$ such that $D1= \varepsilon (D)1$, where 1 is the constant function 1. From this definition we deduce that $\varepsilon({\bf{1}})=1$ (where 1 here is the real number 1), and $\varepsilon \left( \partial_{i_1} \cdots \partial_{i_n} \right)=0$ for $n>0$. It can be verified that $\varepsilon$ indeed satisfies the defining property of a counit as $D = \sum \varepsilon \left (D_{(1)} \right) D_{(2)} = D_{(1)} \varepsilon \left( D_{(2)} \right)$. Finally the antipode is a linear map $S : \mathcal{A}  \mapsto \mathcal{A}$ such that $\sum S\left (D_{(1)} \right) D_{(2)} = D_{(1)} S\left( D_{(2)} \right) = \varepsilon (D) {\bf{1}}$. It can be verified that $S({\bf{1}})={\bf{1}}$ and $S\left( \partial_{i_1} \cdots \partial_{i_n} \right)= (-1)^n \left( \partial_{i_1} \cdots \partial_{i_n} \right)$ for $n>0$.

The Hopf algebra of differential operators can be quite effective in practice. It has for instance been used to describe the hierarchy of Green functions in quantum field theory \cite{brouder2003many}.


\subsection{Trees from derivatives}
In this subsection, we present a correspondence between the structure of the higher order differentials in the Fa\`a di Bruno formula which computes the $n$n-th derivative of the composition $f \circ g$ of two functions $f$ and $g$ in terms of their respective derivatives and the elements of the set of rooted trees. 

The original Fa\`a di Bruno formula expands derivatives of compositions of functions on $\mathbb{R}$ as 

\begin{eqnarray}
\frac{d^n}{dx^n} f\left( g\left( x \right) \right)= \sum_{k=1}^n \sum_{\substack{j_1 + 2 j_2 + \cdots + n j_n = n \\ j_1 +  j_2 + \cdots +  j_n = k }} \frac{n!}{j_1! j_2! \cdots j_n!} f^{(k)}\left( g\left( x \right) \right) \left( \frac{g'(x)}{1!} \right)^{j_1} \left( \frac{g''(x)}{2!} \right)^{j_2} \cdots \left( \frac{g^{(n)}(x)}{n!} \right)^{j_n}.   
\end{eqnarray}

\noindent Its combinatorial nature emerges when expressing the Fa\`a di Bruno formula in terms of the partial Bell polynomials $B_{n,k}$ as

\begin{eqnarray}
\frac{d^n}{dx^n} f\left( g\left( x \right) \right)= \sum_{k=1}^n f^{(k)}\left( g\left( x \right) \right) B_{n,k} \left( g'(x), g''(x), \dots, g^{(n+1-k)}(x) \right). 
\end{eqnarray}

\noindent A rooted tree can be defined as a class of oriented (non planar) graphs with a finite number of vertices, among which is a special one called the root, such that any vertex admits exactly one incoming edge, except the root which has only outgoing edges. The correspondence between the derivatives of the composition of functions and rooted trees comes from the fact that differentiation amounts to creating edges. The process can be described as follows \cite{chan2014relations}. In the set of rooted trees $\mathbb{T}_R$, we attach $f^{(k)}$ to the root from which, upon $k$-th differentiation, will be attached $k$ edges. Connected to these edges are $k$ trees having at most one incoming and also at most one outgoing edge called ladder (linear) trees, which are obtained by deconcatenation of the monomial 

\begin{eqnarray}
\prod_{\substack{j_1 + 2 j_2 + \cdots + n j_n = n \\ j_1 +  j_2 + \cdots +  j_n = k }} \left( g^{(n)} \right)^{J_n}.  
\end{eqnarray}

\noindent The reason for obtaining ladder trees is that while the differentiation of $f^{(k)}\left( g\left( x \right) \right)$ creates several outgoing edges from the root to which $f^{(k)}$ is attached because of the chain rule, only at most one incoming and also at most one outgoing edge will will be created between the vertices attached by $g^{(k)}$, because the chain rule does not apply to the differentiation of $g^{(k)}(x)$. As a result, ladder trees with number of vertices equal to the order of differentiation are produced. Then, considering that $f$ and $g$ are  functions with convergent Taylor series about $x = 0$ which leave the origin invariant, for any $n \in \mathbb{N}$, we have that $f_n=f^{(n)}(0)$ and $g_n=g^{(n)}(0)$. We can therefore establish a correspondence between the expansion of the Fa\`a di Bruno formula and the expansion of the composition of functions. The corresponding set of rooted trees $\mathbb{T}_{lR}$ in $\mathbb{T}_R$ is

\begin{center}
$\mathbb{T}_{lR}$ =  $\Biggl\{$
\begin{forest}
for tree={circle,draw,fill,minimum width=2pt,inner sep=0pt,parent anchor=center,child anchor=center,s sep+=0pt,l sep-=5pt,grow=north,}
[, for tree={l=0} [ ] ] 
\end{forest}~, \hspace{0.5cm}
\begin{forest}
for tree={circle,draw,fill,minimum width=2pt,inner sep=0pt,parent anchor=center,child anchor=center,s sep+=0pt,l sep-=5pt,grow=north,}
[, for tree={l=0} [ [ ] ] ] 
\end{forest}~,
\begin{forest}
for tree={circle,draw,fill,minimum width=2pt,inner sep=0pt,parent anchor=center,child anchor=center,s sep+=0pt,l sep-=5pt,grow=north,}
[, for tree={l=0} [ ] [ ]  ] 
\end{forest}, \hspace{0.5cm}
\begin{forest}
for tree={circle,draw,fill,minimum width=2pt,inner sep=0pt,parent anchor=center,child anchor=center,s sep+=0pt,l sep-=5pt,grow=north,}
[, for tree={l=0} [ [[]] ]  ] 
\end{forest}~,
\begin{forest}
for tree={circle,draw,fill,minimum width=2pt,inner sep=0pt,parent anchor=center,child anchor=center,s sep+=0pt,l sep-=5pt,grow=north,}
[, for tree={l=0} [  ] [[] ] ] 
\end{forest},
\begin{forest}
for tree={circle,draw,fill,minimum width=2pt,inner sep=0pt,parent anchor=center,child anchor=center,s sep+=0pt,l sep-=5pt,grow=north,}
[, for tree={l=0} [ ] [ ] [ ] ] 
\end{forest},  \hspace{0.5cm}
$ \ldots ~~ \Biggl\}$,
\end{center}

\noindent the associated set $\mathcal{F} \left( \mathbb{T}_{lR} \right)$ of respective differential terms giving the derivatives of $f \circ g$ also called elementary differentials in the context of BUtcher series \cite{chan2014relations}, is

\begin{eqnarray}
\mathcal{F} \left( \mathbb{T}_{lR} \right) = \{ f'g', \hspace{0.5cm} f'g'', f''(g',g'), \hspace{0.5cm} f'g''', f''(g'',g'), f'''(g',g',g'), \hspace{0.5cm} \ldots   \},    
\end{eqnarray}

\noindent and the corresponding set $\mathcal{F} \left( \mathbb{T}_{lR} \right)\left( 0 \right)$ of respective terms in the expansion of $f \left( g(x) \right)$ is

\begin{eqnarray}
\mathcal{F} \left( \mathbb{T}_{lR} \right) \left( 0 \right) = \{ f_1, \hspace{0.5cm} f_1g_2, f_2g_1^2, \hspace{0.5cm} f_1g_3, f_2g_1g_2, f_3g_1^3, \hspace{0.5cm} \ldots   \}.    
\end{eqnarray}

\noindent The derivation of the log partition function as a sum over rooted trees is made more explicit below, from the combinatorics of renormalization in quantum field theory.

\subsection{Feynman graphs and Hopf (sub)algebra of rooted trees}
Graphical techniques using trees have been used in many algebraic constructions leading to important developments in various fields. In high energy physics particularly, a groundbreaking achievement in connection with the process of renormalization in perturbative quantum field theory was reached when Kreimer realized that the underlying structure of BPHZ renormalization is captured by a Hopf algebra of combinatorial nature \cite{Kreimer:1997dp}. This accomplishment opened the way to a fertile interaction between mathematics and physics, marked by the seminal work of Connes and Kreimer, who introduced the Connes-Kreimer Hopf algebra of rooted trees thus giving a sturdy algebraic framework for the BPHZ renormalization \cite{Connes:1998qv}.

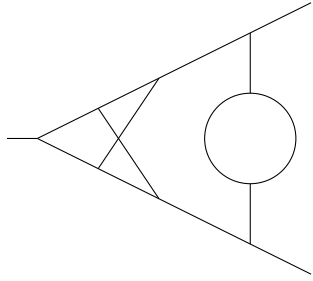
\begin{figure}[H]
\centering
\begin{tikzpicture}[scale=0.4]
\draw[-] (-6,0)--(-5,0);
\draw[](2,0) circle (1.5);
\draw[-] (2,1.5)--(2,3.5);
\draw[-] (2,-1.5)--(2,-3.5);
\draw[-] (-5,0)--(4,4.5);
\draw[-] (-5,0)--(4,-4.5);
\draw[-] (-3,1)--(-1,-2);
\draw[-] (-1,2)--(-3,-1);
\end{tikzpicture}    
\caption{A Feynman graph} 
\label{fig1}
\end{figure}

The intricacy of renormalization can be considered in the following way \cite{Krajewski:1998xi,brouder2004trees}: an integral is attached to certain graphs called Feynman graphs, according to certain rules called the Feynman rules (see Fig. \ref{fig1}).

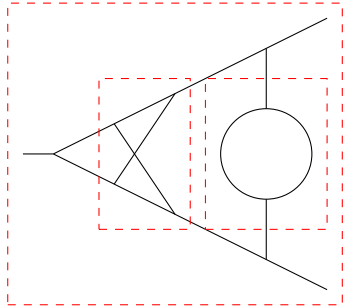
\begin{figure}[H]
\centering
\begin{tikzpicture}[scale=0.4]
\draw[-] (-6,0)--(-5,0);
\draw[](2,0) circle (1.5);
\draw[-] (2,1.5)--(2,3.5);
\draw[-] (2,-1.5)--(2,-3.5);
\draw[-] (-5,0)--(4,4.5);
\draw[-] (-5,0)--(4,-4.5);
\draw[-] (-3,1)--(-1,-2);
\draw[-] (-1,2)--(-3,-1);
\draw[dashed,red] (-0,2.5) rectangle (4,-2.5);
\draw[dashed,red] (-3.5,2.5) rectangle (-0.5,-2.5);
\draw[dashed,red] (-6.5,5) rectangle (4.5,-5);
\end{tikzpicture}    
\caption{Subdivergences of the Feynman graph} 
\label{fig2}
\end{figure}

\noindent These integrals turn out to be divergent, because of the presence of loops in the Feynman graphs; each loop creates a subdivergence in the associated integral (see Fig. \ref{fig2}).

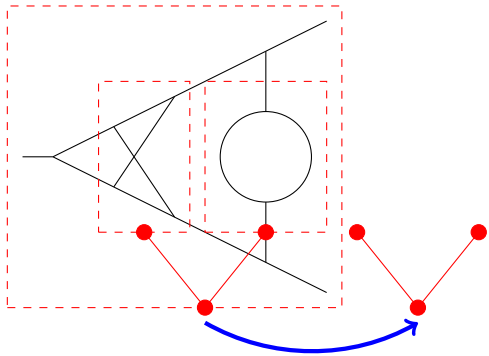
\begin{figure}[H]
\centering
\begin{tikzpicture}[scale=0.4]
\draw[-] (-6,0)--(-5,0);
\draw[](2,0) circle (1.5);
\draw[-] (2,1.5)--(2,3.5);
\draw[-] (2,-1.5)--(2,-3.5);
\draw[-] (-5,0)--(4,4.5);
\draw[-] (-5,0)--(4,-4.5);
\draw[-] (-3,1)--(-1,-2);
\draw[-] (-1,2)--(-3,-1);
\draw[dashed,red] (-0,2.5) rectangle (4,-2.5);
\draw[dashed,red] (-3.5,2.5) rectangle (-0.5,-2.5);
\draw[dashed,red] (-6.5,5) rectangle (4.5,-5);
\draw[-,red] (-0,-5) -- (-2,-2.5);
\draw[-,red] (-0,-5) -- (2,-2.5);
\filldraw[red] (0,-5) circle (0.25);
\filldraw[red] (-2,-2.5) circle (0.25);
\filldraw[red] (2,-2.5) circle (0.25);
\draw[ultra thick,blue, ->] (0,-5.5) arc (240:300:7.0);
\draw[-,red] (7,-5) -- (5,-2.5);
\draw[-,red] (7,-5) -- (9,-2.5);
\filldraw[red] (7,-5) circle (0.25);
\filldraw[red] (5,-2.5) circle (0.25);
\filldraw[red] (9,-2.5) circle (0.25);
\end{tikzpicture}   
\caption{Rooted tree associated to the Feynman graph} 
\label{fig3}
\end{figure}

\noindent The renormalization procedure \cite{Collins:1984xc} is used to make sense of these integrals. In order to renormalize the Feynman graphs associated to integrals, one must remove the subdivergences in a complicated way known as Zimmermann's forest formula \cite{zeidler2008quantum}. In 1998, Kreimer discovered \cite{Kreimer:1997dp} that this formula could be understood as a Hopf algebra over rooted trees. Then, in the Connes-Kreimer algebraic setting, the renormalization consists in associating to each Feynman graph a rooted tree that describes the structure of the subdivergences of the graph (see Fig. \ref{fig3}).

\noindent The Connes-Kreimer formalism treats disjoint, nested and overlapping divergences on the same footing. Indeed, while the assignment of a unique rooted tree to a Feynman graph with subdivergences occurs if and only if all subdivergences of the Feynman graph are either disjoint or nested, the case of overlapping subdivergences also follows the combinatorics governed by the Hopf algebra of rooted trees, and in such case, a Feynman graph corresponds to a linear combination (a sum) of rooted trees, i.e a forest.
 
\noindent After a regularization step, the Feynman rules induce an algebra morphism from the algebra of rooted trees to the algebra of functions. Through algebra morphism, it is eventually possible to express the log partition function as a sum over rooted trees.

\subsubsection{Hopf algebra of rooted trees}
We introduce fundamental definitions concerning rooted trees and their (Connes-Kreimer) Hopf algebra. A (non planar) rooted tree is either the empty set, or a finite connected oriented graph without loops in which every vertex has exactly one incoming edge, except for a distinguished vertex (the root) which has no incoming edge, but only outgoing edges. The set of edges and vertices of a rooted tree $T$ is denoted $E(T)$ and $V(T)$, respectively. Let $\mathbb{T}_R$  be the set of (isomorphism classes of) rooted trees. We list below all elements of this set up to degree 4

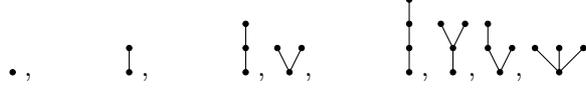
\begin{center}
\begin{forest}
for tree={circle,draw,fill,minimum width=2pt,inner sep=0pt,parent anchor=center,child anchor=center,s sep+=0pt,l sep-=5pt,grow=north,}
[, for tree={l=0} ] 
\end{forest}~, \hspace{1cm}
\begin{forest}
for tree={circle,draw,fill,minimum width=2pt,inner sep=0pt,parent anchor=center,child anchor=center,s sep+=0pt,l sep-=5pt,grow=north,}
[, for tree={l=0} [ ] ] 
\end{forest}~, \hspace{1cm}
\begin{forest}
for tree={circle,draw,fill,minimum width=2pt,inner sep=0pt,parent anchor=center,child anchor=center,s sep+=0pt,l sep-=5pt,grow=north,}
[, for tree={l=0} [ [ ] ] ] 
\end{forest}~,
\begin{forest}
for tree={circle,draw,fill,minimum width=2pt,inner sep=0pt,parent anchor=center,child anchor=center,s sep+=0pt,l sep-=5pt,grow=north,}
[, for tree={l=0} [ ] [ ]  ] 
\end{forest}, \hspace{1cm}
\begin{forest}
for tree={circle,draw,fill,minimum width=2pt,inner sep=0pt,parent anchor=center,child anchor=center,s sep+=0pt,l sep-=5pt,grow=north,}
[, for tree={l=0} [ [[]] ]  ] 
\end{forest}~,
\begin{forest}
for tree={circle,draw,fill,minimum width=2pt,inner sep=0pt,parent anchor=center,child anchor=center,s sep+=0pt,l sep-=5pt,grow=north,}
[, for tree={l=0} [ [] [] ] ] 
\end{forest},
\begin{forest}
for tree={circle,draw,fill,minimum width=2pt,inner sep=0pt,parent anchor=center,child anchor=center,s sep+=0pt,l sep-=5pt,grow=north,}
[, for tree={l=0} [  ] [[] ] ] 
\end{forest},
\begin{forest}
for tree={circle,draw,fill,minimum width=2pt,inner sep=0pt,parent anchor=center,child anchor=center,s sep+=0pt,l sep-=5pt,grow=north,}
[, for tree={l=0} [ ] [ ] [ ] ] 
\end{forest}   
\end{center}

\noindent The commutative, unital, associative $\mathbb{C}$-algebra of rooted trees $\mathcal{A}_R$ is the polynomial algebra generated by the symbols $T$, one for each isomorphism class of rooted trees. The unit denoted by 1 is the empty tree. The grading of $\mathcal{A}_R$ is defined in terms of the number of vertices of $\# (T) = \vert V(T) \vert$, which is extended to monomials (i.e products of rooted trees) also called rooted forests, by $\# ( T_1T_2 \cdots T_n )= \sum_(i=1)^n \# (T_i)$, turning $\mathcal{A}_R = \bigoplus_{n \geq 0} \mathcal{A}_R^{(n)} $ into a graded, connected, unital, commutative, associative $\mathbb{C}$-algebra. Then the Connes-Kreimer Hopf algebra $\mathcal{H}_R= \left( \mathcal{A}_R, \Delta, \epsilon  \right)$ is the algebra $\mathcal{A}_R$ endowed with the counit $\epsilon: \mathcal{H}_R \mapsto \mathbb{C}$ defined by $\epsilon (1) := 1_{\mathbb{C}}$ and $\epsilon ( T_1T_2 \cdots T_n ) = 0$ if $T_1T_2 \cdots T_n$ are trees, as well as the coproduct $\mathcal{H}_R \mapsto \mathcal{H}_R \otimes \mathcal{H}_R$ defined in terms of admissible cuts on the rooted trees as follows \cite{calaque2011two}. First consider a rooted forest $F$, and impose a path from a root to $y$ passing through $x$. The set $V(F)$ of vertices of the forest $F$ is then endowed with a partial order defined by $x \leq y$. Any subset $B$ of $V(F)$ defines a subforest $F \vert _B$ of $F$ by keeping the edges of $F$  which link two elements of $B$. The structure of the partial order on sets is given by restriction of the partial order to $W$, and the minimal elements are the roots of the subforest. The coproduct is then defined by 

\begin{eqnarray}
\Delta (F)= \sum_{\substack{A \sqcup B = V(F) \\ A<B}}  F \vert _A \otimes F \vert _B. 
\end{eqnarray}

\noindent A couple $(A,B)$ is also called an admissible cut, with crown (or pruning) $F \vert _A$, and trunk $F \vert _B$. In \cite{Connes:1998qv}, the coproduct of $\mathcal{H}_R$ was given as

\begin{eqnarray}
\Delta 1:= 1 \otimes 1, \hspace{0.5cm} \Delta T= T \otimes 1 + 1 \otimes T + \sum_{c \in C(T)} P_c(T) \otimes R_c(T), 
\end{eqnarray}

\noindent where $C(T)$ is the list of admissible cuts, $P_c(T)$ is the pruning, i.e the subforest formed by the edges above the cut $c$, and $R_c(T)$ is the subforest formed by the edges under $c$. For instance

\begin{eqnarray}
\Delta \left( 
\begin{forest}
for tree={circle,draw,fill,minimum width=2pt,inner sep=0pt,parent anchor=center,child anchor=center,s sep+=0pt,l sep-=5pt,grow=north,}
[, for tree={l=0} [ ] ] 
\end{forest}
\right) &=& 
\begin{forest}
for tree={circle,draw,fill,minimum width=2pt,inner sep=0pt,parent anchor=center,child anchor=center,s sep+=0pt,l sep-=5pt,grow=north,}
[, for tree={l=0} [ ] ] 
\end{forest}
\otimes 1 + 1 \otimes 
\begin{forest}
for tree={circle,draw,fill,minimum width=2pt,inner sep=0pt,parent anchor=center,child anchor=center,s sep+=0pt,l sep-=5pt,grow=north,}
[, for tree={l=0} [ ] ] 
\end{forest} 
+ 
\begin{forest}
for tree={circle,draw,fill,minimum width=2pt,inner sep=0pt,parent anchor=center,child anchor=center,s sep+=0pt,l sep-=5pt,grow=north,}
[, for tree={l=0} ] 
\end{forest}
\otimes 
\begin{forest}
for tree={circle,draw,fill,minimum width=2pt,inner sep=0pt,parent anchor=center,child anchor=center,s sep+=0pt,l sep-=5pt,grow=north,}
[, for tree={l=0} ] 
\end{forest} 
\\
\Delta \left(
\begin{forest}
for tree={circle,draw,fill,minimum width=2pt,inner sep=0pt,parent anchor=center,child anchor=center,s sep+=0pt,l sep-=5pt,grow=north,}
[, for tree={l=0} [ ] [ ] ] 
\end{forest}
\right) &=& 
\begin{forest}
for tree={circle,draw,fill,minimum width=2pt,inner sep=0pt,parent anchor=center,child anchor=center,s sep+=0pt,l sep-=5pt,grow=north,}
[, for tree={l=0} [ ] [ ] ] 
\end{forest}
\otimes 1 +1 \otimes 
\begin{forest}
for tree={circle,draw,fill,minimum width=2pt,inner sep=0pt,parent anchor=center,child anchor=center,s sep+=0pt,l sep-=5pt,grow=north,}
[, for tree={l=0} [ ] [ ] ] 
\end{forest} 
+2 ~
\begin{forest}
for tree={circle,draw,fill,minimum width=2pt,inner sep=0pt,parent anchor=center,child anchor=center,s sep+=0pt,l sep-=5pt,grow=north,}
[, for tree={l=0}  ] 
\end{forest}
\otimes 
\begin{forest}
for tree={circle,draw,fill,minimum width=2pt,inner sep=0pt,parent anchor=center,child anchor=center,s sep+=0pt,l sep-=5pt,grow=north,}
[, for tree={l=0} [ ] ] 
\end{forest}
+
\begin{forest}
for tree={circle,draw,fill,minimum width=2pt,inner sep=0pt,parent anchor=center,child anchor=center,s sep+=0pt,l sep-=5pt,grow=north,}
[, for tree={l=0} ] 
\end{forest} ~
\begin{forest}
for tree={circle,draw,fill,minimum width=2pt,inner sep=0pt,parent anchor=center,child anchor=center,s sep+=0pt,l sep-=5pt,grow=north,}
[, for tree={l=0} ] 
\end{forest}
\otimes 
\begin{forest}
for tree={circle,draw,fill,minimum width=2pt,inner sep=0pt,parent anchor=center,child anchor=center,s sep+=0pt,l sep-=5pt,grow=north,}
[, for tree={l=0} ] 
\end{forest}
\end{eqnarray}

\noindent The linear operator $B_+: \mathcal{H}_R \mapsto \mathcal{H}_R$ also known as the grafting operator is a map that takes any forest to a tree, by connecting the roots in the monomials of rooted trees making the forest to a new adjoined root. For example

\begin{eqnarray}
B_+ \left(
\begin{forest}
for tree={circle,draw,fill,minimum width=2pt,inner sep=0pt,parent anchor=center,child anchor=center,s sep+=0pt,l sep-=5pt,grow=north,}
[, for tree={l=0} ] 
\end{forest}
~
\begin{forest}
for tree={circle,draw,fill,minimum width=2pt,inner sep=0pt,parent anchor=center,child anchor=center,s sep+=0pt,l sep-=5pt,grow=north,}
[, for tree={l=0} [ ] ] 
\end{forest}
\right)
=
\begin{forest}
for tree={circle,draw,fill,minimum width=2pt,inner sep=0pt,parent anchor=center,child anchor=center,s sep+=0pt,l sep-=5pt,grow=north,}
[, for tree={l=0} [ [] ] [ ] ] 
\end{forest}
\end{eqnarray}

\subsubsection{The ladder tree Hopf algebra}
The Connes-Kreimer Hopf algebra of rooted trees contains a subalgebra called the Hopf algebra of rooted ladder trees denoted $\mathcal{H}_L$ \cite{Figueroa:2004hb,Ebrahimi-Fard:2004twg,Chryssomalakos:2001vh}, which is generated by rooted ladder trees

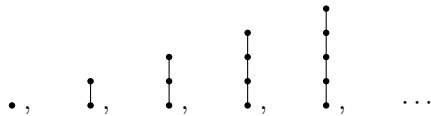
\begin{center}
\begin{forest}
for tree={circle,draw,fill,minimum width=2pt,inner sep=0pt,parent anchor=center,child anchor=center,s sep+=0pt,l sep-=5pt,grow=north,}
[, for tree={l=0} ] 
\end{forest}~,
\hspace{0.5cm}
\begin{forest}
for tree={circle,draw,fill,minimum width=2pt,inner sep=0pt,parent anchor=center,child anchor=center,s sep+=0pt,l sep-=5pt,grow=north,}
[, for tree={l=0} [ ] ] 
\end{forest}~,
\hspace{0.5cm}
\begin{forest}
for tree={circle,draw,fill,minimum width=2pt,inner sep=0pt,parent anchor=center,child anchor=center,s sep+=0pt,l sep-=5pt,grow=north,}
[, for tree={l=0} [ [ ] ] ] 
\end{forest}~,
\hspace{0.5cm}
\begin{forest}
for tree={circle,draw,fill,minimum width=2pt,inner sep=0pt,parent anchor=center,child anchor=center,s sep+=0pt,l sep-=5pt,grow=north,}
[, for tree={l=0} [ [ [ ] ] ] ] 
\end{forest}~,
\hspace{0.5cm}
\begin{forest}
for tree={circle,draw,fill,minimum width=2pt,inner sep=0pt,parent anchor=center,child anchor=center,s sep+=0pt,l sep-=5pt,grow=north,}
[, for tree={l=0} [ [ [ [ ] ] ] ] ] 
\end{forest}~,
\hspace{0.5cm} $\cdots$
\end{center}

\noindent Denoting the ladder trees with $n$ vertices in the above type of linearly ordered set by $l_n \in \mathcal{H}_L \subset \mathcal{H}_R$, the coproduct of $\mathcal{H}_L$ then becomes 

\begin{eqnarray}
\Delta ( l_n) = \sum_{i=0}^{n} l_i \otimes l_{n-i}.
\end{eqnarray}

\noindent This is a commutative, cocommutative Hopf algebra, isomorphic to the Hopf algebra of symmetric functions.

Having established the necessary background, we show in the next section how the log partition function generates rooted trees indexed by Hurwitz numbers.

\subsection{Log partition function as a sum over rooted trees indexed by Hurwitz numbers}
A Hurwitz number counts the number of non-equivalent branched coverings of a surface with a prescribed set of branch points and branched profile. Although branched coverings first appeared in \cite{riemann1857theorie}, their enumeration was studied in a systematic way by Hurwitz who observed that the counting of branched coverings could be interpreted in terms of permutation factorizations  \cite{hurwitz1891riemann,hurwitz1901ueber}. Ever since, Hurwitz numbers have been an important subject in mathematics and physics, with an enormous amount of literature dedicated to them \cite{frobenius1896gruppencharaktere,frobenius1906reellen,mednykh1986number,jones1995enumeration,natanzon2010simple,Alexeevski:2007js,Alexeevski:2007js,Mironov:2009cj,Mironov:2010yg}. They have been found notably in the context of string theory after a crucial observation made in \cite{dijkgraaf1995mirror,dijkgraaf1989geometrical} from which many works followed, in integrable systems with early works \cite{Okounkov:2000fna,Okounkov:2002cja}, or in matrix models \cite{Chekhov:2005kd,deMelloKoch:2010hav,Alexandrov:2010th,Orlov:2017nub,Orlov:2018mbz}.

Recalling the definition of Hurwitz numbers \cite{cavalieri2016riemann}, let $Y$ be a connected Riemann surface of genus $g$. Define the set $B = \{ y_1, \ldots, y_d \} \in Y $, and let $\lambda_1, \ldots, \lambda_d$ be partitions of the positive integer $n$. Then the Hurwitz number can be defined as the sum

\begin{eqnarray}
H_{X \xrightarrow[]{n} Y} \left( \lambda_1, \ldots, \lambda_d \right) = \sum_{\left|f \right|} \frac{1}{\left| \rm{Aut} (f) \right|}
\end{eqnarray}

\noindent that runs over each isomorphism class of $f: X \mapsto Y$ where

\begin{enumerate}
    \item $f$ is a holomorphic map of Riemann surfaces;
    \item $X$ is connected and has genus $h$;
    \item the branch locus of $f$ is $B = \{ y_1, \ldots, y_d \}$;
    \item the ramification profile of $f$ at $y_i$ is $\lambda_i$.
\end{enumerate}

\noindent Hurwitz numbers arise in two different flavors, depending on whether the covering space $X$ of $Y$ is connected or not. Although we will start with the above definition of the connected Hurwitz number, our focus is on the disconnected theory, so we also give the general formula for disconnected Hurwitz numbers, mentioning beforehand that we will restrict our attention to the target space with genus $g=0$. The problem at hand is then attacked by using the representation theory of the symmetric group. 

Let $\lambda_1, \ldots, \lambda_d$ be partitions of the positive integer $n$. Recall from the representation theory of the symmetric group that $\mathfrak{Z} \left( \mathbb{C} \left[ S_n \right] \right)$ is a vector space with dimension equal to the number of partitions of $n$ and basis indexed by conjugacy classes of permutations. Denoting the basis element associated to the corresponding conjugacy class by $C_{\lambda_i}$ for every $i \in \left[ 1;d \right]$, the genus zero disconnected Hurwitz number takes the form

\begin{eqnarray}
\label{disconnect HN}
H^{\bullet}_{X \xrightarrow[]{n} 0} \left( \lambda_1, \ldots, \lambda_d \right) = \frac{1}{n!} \left[ C_e \right] C_{\lambda_d} \cdots C_{\lambda_{2}} C_{\lambda_{1}},
\end{eqnarray}

\noindent where $\left[ C_e \right] C_{\lambda_d} \cdots C_{\lambda_{2}} C_{\lambda_{1}}$ is the coefficient of $C_{e} = \{ e \}$ after writing the product as a linear combination of the basis element $C_{\lambda_i}$.

After restricting the genus of the base and target Riemann surfaces to be zero, we further impose $d=2$ and $\lambda_1 = \lambda_2 = (n)$. The expression of connected Hurwitz numbers becomes

\begin{eqnarray}
\label{connected HN}
H_{0 \xrightarrow[]{n} 0} \left( (n), (n) \right) = \frac{1}{n}.
\end{eqnarray}

 As a subalgebra of the Hopf algebra of rooted trees $\mathcal{H}_R$, the ladder tree Hopf algebra $\mathcal{H}_L$ is also known to be isomorphic to a Hopf algebra of polynomials \cite{Figueroa:2004hb}, and a correspondence can be established between the polynomial coordinates $\mathcal{G}_1,  \ldots, \mathcal{G}_n$ and the generators of $\mathcal{H}_L$ which we will denote $l_1, \ldots, l_n$, obtained by making use of iterated action of the operator $B_+$ on the empty tree $\mathbbm{1}$, the first three of which read 

\begin{center}
$l_1 = B_+ \left( \mathbbm{1} \right) =$ \begin{forest}
for tree={circle,draw,fill,minimum width=2pt,inner sep=0pt,parent anchor=center,child anchor=center,s sep+=0pt,l sep-=5pt,grow=north,}
[, for tree={l=0} ] 
\end{forest}~,
\hspace{0.5cm}
$l_2 = B_+ \left(  B_+ \left( \mathbbm{1} \right) \right) =$ \begin{forest}
for tree={circle,draw,fill,minimum width=2pt,inner sep=0pt,parent anchor=center,child anchor=center,s sep+=0pt,l sep-=5pt,grow=north,}
[, for tree={l=0} [ ] ] 
\end{forest}~,
\hspace{0.5cm}
$l_3 = B_+ \left( B_+ \left(  B_+ \left( \mathbbm{1} \right) \right) \right) =$ \begin{forest}
for tree={circle,draw,fill,minimum width=2pt,inner sep=0pt,parent anchor=center,child anchor=center,s sep+=0pt,l sep-=5pt,grow=north,}
[, for tree={l=0} [ [ ] ] ] 
\end{forest}~.
\end{center}

\noindent This implies that the sum $F \left( \mathcal{G}_1, \ldots, \mathcal{G}_n \right)$ that is subjected to exponentiation in Eq. (\ref{PE}) can be expressed as a sum of ladder rooted trees 

\begin{eqnarray}
\label{sum of ladder trees}
F \left( l_1, \ldots, l_n \right) =\sum_{n=1}^{\infty} \frac{1}{n} l_n = \hspace{0.25cm}
\begin{forest}
for tree={circle,draw,fill,minimum width=2pt,inner sep=0pt,parent anchor=center,child anchor=center,s sep+=0pt,l sep-=5pt,grow=north,}
[, for tree={l=0} ] 
\end{forest} \hspace{0.25cm} + \hspace{0.25cm} 
\frac{1}{2} \hspace{0.1cm}
\begin{forest}
for tree={circle,draw,fill,minimum width=2pt,inner sep=0pt,parent anchor=center,child anchor=center,s sep+=0pt,l sep-=5pt,grow=north,}
[, for tree={l=0} [ ] ] 
\end{forest} \hspace{0.25cm} + \hspace{0.25cm}
\frac{1}{3} \hspace{0.1cm}
\begin{forest}
for tree={circle,draw,fill,minimum width=2pt,inner sep=0pt,parent anchor=center,child anchor=center,s sep+=0pt,l sep-=5pt,grow=north,}
[, for tree={l=0} [ [ ] ] ] 
\end{forest} \hspace{0.25cm} + \hspace{0.25cm} 
\frac{1}{4} \hspace{0.1cm}
\begin{forest}
for tree={circle,draw,fill,minimum width=2pt,inner sep=0pt,parent anchor=center,child anchor=center,s sep+=0pt,l sep-=5pt,grow=north,}
[, for tree={l=0} [ [ [ ] ] ] ] 
\end{forest} \hspace{0.25cm} + \hspace{0.25cm} 
\frac{1}{5} \hspace{0.1cm}
\begin{forest}
for tree={circle,draw,fill,minimum width=2pt,inner sep=0pt,parent anchor=center,child anchor=center,s sep+=0pt,l sep-=5pt,grow=north,}
[, for tree={l=0} [ [ [ [ ] ] ] ] ] 
\end{forest} \hspace{0.25cm} + \hspace{0.25cm} \cdots.
\end{eqnarray}

\noindent From Eq. (\ref{connected HN}), the first part of our derivation is

\begin{eqnarray}
\boxed{F \left( l_1, \ldots, l_n \right) =\sum_{n=1}^{\infty} \left[H_{0 \xrightarrow[]{n} 0} \left( (n), (n) \right) \right] l_n},
\end{eqnarray}

\noindent where we see that the connected Hurwitz numbers $H_{0 \xrightarrow[]{n} 0} \left( (n), (n) \right)$ are the coefficients of the generators of the Hopf algebra of ladder rooted trees $\mathcal{H}_L$.

\noindent Because connected and disconnected Hurwitz generating functions are related by exponentiation, disconnected Hurwitz numbers appear in the $\left( q^2 \right)$ parameter-inserted version of the exponentiation of the function $F \left( l_1, \ldots, l_n \right)$ that yields $Z_{log}$ as a generating function of rooted trees. The derivation is as follows.

Let $n=1,2, \ldots$ and $k=1, \ldots, n$. Moreover, let $p(n,k)$ denote the tuple of nonnegative integer solutions $j := j_1, \ldots, j_n$ of the system

\begin{eqnarray}
\label{system}
\left\{
\begin{array}{ll}
j_1 + j_2 + \cdots + j_n = k,\\
j_1 + 2 j_2 + \cdots + n j_n = n.
\end{array}
\right.
\end{eqnarray}

\noindent The generalized binomial coefficient defined as 

\begin{eqnarray}
\begin{pmatrix}
n \\
j_1, \ldots, j_k \\
\end{pmatrix}
= \frac{n!}{j_1! j_2! \cdots j_n! \cdot (1!)^{j_1} (2!)^{j_2} \cdots (n!)^{j_n}},
\end{eqnarray}

\noindent can be interpreted in terms of partitions by considering partitions of the set $\{ 1,2, \ldots, n \}$ into $k$ blocks of $j_i$ $i$ elements subsets such that the system (\ref{system}) holds. Then the number of all partitions of this type is equal to the binomial coefficient. Then the partition function $Z_{log}$ becomes

\begin{subequations}
\begin{align}
Z_{log} \left( l_1, \ldots, l_n \right) &= 1 + \sum_{n=1}^{\infty} \frac{1}{n!} \left( \sum_{\substack{\sum_{k=1}^n k j_k =n \\ n \geq 1\\j_k \geq 0 }} \frac{n!}{\prod_{k=1}^n j_k! (k) ^{j_k}}B_+ \left( \prod_{k=1}^n l_k^{j_k} \right) \right) \left( q^2 \right)^n  \\
&= 1 + \sum_{n=1}^{\infty}  \left( \sum_{\substack{\sum_{k=1}^n k j_k =n \\ n \geq 1\\j_k \geq 0 }} \frac{1}{\prod_{k=1}^n j_k! (k) ^{j_k}} B_+ \left( \prod_{k=1}^n l_k^{j_k} \right) \right) \left( q^2 \right)^n  
\end{align}
\end{subequations}

\noindent with its final expression as

\begin{eqnarray}
\label{Hurwitz partition function}
\boxed{Z_{log} \left( l_1, \ldots, l_n \right) =  1 + \sum_{n=1}^{\infty}  \left( \sum_{\substack{\sum_{k=1}^n k j_k =n \\ n \geq 1\\j_k \geq 0 }} \left[ H^{\bullet}_{0 \xrightarrow[]{n} 0} \left( \left( [1]^{j_1}, [2]^{j_2}, \ldots \right),  \left( [1]^{j_1}, [2]^{j_2}, \ldots \right) \right) \right] B_+ \left( \prod_{k=1}^n l_k^{j_k} \right) \right) \left( q^2 \right)^n},
\end{eqnarray}

\noindent and the disconnected Hurwitz numbers expressed as 

\begin{eqnarray}
\label{disconnected HN}
H^{\bullet}_{0 \xrightarrow[]{n} 0} \left( \left( [1]^{j_1}, [2]^{j_2}, \ldots \right),  \left( [1]^{j_1}, [2]^{j_2}, \ldots \right) \right) = \prod_{k=1}^n \frac{1}{j_k!  (k)^{j_k}},
\end{eqnarray}

\noindent where the $\left( [1]^{j_1}, \ldots, [k]^{j_k} \right)$ associated to the trees $B_+ \left( \prod_{k=1}^n l_k^{j_k} \right)$ are such that $[k]^{j_k}= \overbrace{k, \ldots. k}^{j_k ~ \rm{times}}$.

As an example, the set $\{ 1,2,3 \}$ has five partitions. Three of these have two blocks, namely $\{1,2\} \{3\}$, $\{1,3\} \{2\}$ and $\{2,3\} \{1\}$. With data $k=2, j_1 = j_2 = 1$, we associate the rooted tree

\begin{center}
\begin{forest}
for tree={circle,draw,fill,minimum width=2pt,inner sep=0pt,parent anchor=center,child anchor=center,s sep+=0pt,l sep-=5pt,grow=north,}
[, for tree={l=0} [ [] ] [ ] ] 
\end{forest}    
\end{center}

\noindent to each of them. Then the Hurwitz number can be computed as

\begin{eqnarray}
H^{\bullet}_{0 \xrightarrow[]{3} 0} \left( \left( 1,2 \right) , \left( 1,2 \right) \right) = \frac{1}{1! 1! (1)^1 (2)^1} = \frac{1}{2}.
\end{eqnarray}

\noindent This result can be obtained using Eq. (\ref{disconnect HN}) known in the literature, by considering the basis element $C_{(1,2)} = (12) + (13) + (23)$ of the class algebra $\mathfrak{Z} \left( \mathbb{C} \left[ S_3 \right] \right)$. Then

\begin{eqnarray}
\left[ (12) + (13) + (23) \right]  \left[ (12) + (13) + (23) \right] = 3 e + 3 (123) + 3 (132)= 3 C_e + 3 C_{(3)},
\end{eqnarray}

\noindent and

\begin{eqnarray}
H^{\bullet}_{0 \xrightarrow[]{3} 0} \left( \left( 1,2 \right) , \left( 1,2 \right) \right) = \frac{1}{3!} \cdot 3 = \frac{1}{2}. 
\end{eqnarray}

\noindent Similarly, one of the five partitions of the set $\{ 1,2,3 \}$ has three blocks, namely $\{1\} \{2\} \{3\}$. With data $k=3, j_1=3,j_2=j_3=0$, we associate to the three-block partition the rooted tree 

\begin{center}
\begin{forest}
for tree={circle,draw,fill,minimum width=2pt,inner sep=0pt,parent anchor=center,child anchor=center,s sep+=0pt,l sep-=5pt,grow=north,}
[, for tree={l=0} [] [] [] ] 
\end{forest},    
\end{center}

\noindent and the corresponding Hurwitz number is computed as

\begin{eqnarray}
H^{\bullet}_{0 \xrightarrow[]{3} 0} \left( \left( 1,1,1 \right) , \left( 1,1,1 \right) \right) = \frac{1}{3!} = \frac{1}{6}.
\end{eqnarray}

\noindent Finally, one of the five partitions of the set $\{ 1,2,3 \}$ has one block, namely $\{ 1,2,3 \}$ itself. With data $k=1, j_1=j_2=0, j_3=1$, we associate to the one-block partition the rooted tree 

\begin{center}
\begin{forest}
for tree={circle,draw,fill,minimum width=2pt,inner sep=0pt,parent anchor=center,child anchor=center,s sep+=0pt,l sep-=5pt,grow=north,}
[, for tree={l=0} [[[]]]  ] 
\end{forest},    
\end{center}

\noindent and the corresponding Hurwitz number is computed as

\begin{eqnarray}
H^{\bullet}_{0 \xrightarrow[]{3} 0} \left( \left( 3 \right) , \left( 3 \right) \right) = \frac{1}{3}.
\end{eqnarray}

\noindent The last case illustrates the general fact that when $k=1$, i.e for one block partitions,

\begin{eqnarray}
H^{\bullet}_{0 \xrightarrow[]{n} 0} \left( \left( n \right) , \left( n \right) \right) = H_{0 \xrightarrow[]{n} 0} \left( \left( n \right) , \left( n \right) \right). 
\end{eqnarray}

\subsection{From covering maps to maps of rooted trees, counting permutations}
In \cite{hoffman2001analogue}, a theory of universal covers for posets was developed. In particular,considering a partially ordered set $P$ to be the poset of rooted trees, the map $\pi:P \mapsto P$ from the universal cover of $P$ denoted $\tilde{P}$ to P was developed, such that the rank-$n$ elements of $\tilde{P}$ are permutations $\rho=s_1s_22\cdots s_n$ of $\left\{1,2,...,n\right\}$ associated with labelled rooted trees, and the map $\pi \left( \rho \right)$ is just the rooted tree obtained by forgetting the labels. In the theoretical physics literature, on one hand, holomorphic covering maps have been associated to Feynman diagrams in proposals of worldsheet duals for AdS spaces, where string worldsheets corresponding to covering maps are related to gauge theory Feynman diagrams through the Strebel parametrization of the moduli space of Riemann surfaces, which allows an interpretation of Feynman diagrams in terms of moduli spaces of Riemann surfaces (see \cite{Gopakumar:2005fx} for early work on the subject). On the other hand, simple Hurwitz spaces have also been defined as the space of holomorphic maps from worldsheet to target space \cite{deMelloKoch:2010hav}, and it has been shown that the Riemann surfaces appearing as covering spaces, and equivalently the Feynman diagrams corresponding to Hurwitz classes consist of string worldsheets. This shows that Hurwitz numbers have a very natural interpretation in terms of a string worldsheet. The derivation of $Z_{log}$ as a sum of rooted trees presented in the previous section is inspired by these works to show that the Hurwitz numbers generated by $Z_{log}$ not only count Riemann surfaces but also enumerate maps of rooted trees. In fact, regardless of these different objects, what are really being counted are permutations. Below we take a look at permutations on trees, as they apply to the counting generated by $Z_{log}$.

From the tree-level derivation of $Z_{log}$, we see that the trees of the log partition function are composed of ladder trees whose roots are connected to an added common root. This turns the ladder trees into subtrees of the trees with an added root. The next step is to introduce symmetry groups on the trees. We then introduce the group of permutations of $\left\{ 1,2, \ldots, n \right\}$ in the vertices of the ladder trees, by assigning a label between 1 and $n$ to each vertex, and apply permutations on the labels, such that each ladder tree carries a cycle decomposition of the permutations. We only consider isomorphism classes of rooted trees, and choose a representation in each isomorphism class. As an example, the trees

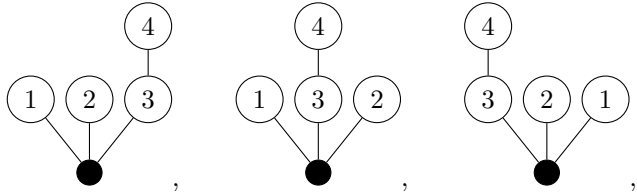
\begin{center}
\begin{forest}
for tree={l sep=0.2em, s sep=0.4em, anchor=center,grow=north}
[,circle,draw,fill [$3$,circle,draw [$4$,circle,draw] ] [$2$,circle,draw ] [$1$,circle,draw]]  
\end{forest}, \hspace{0.5cm} 
\begin{forest}
for tree={l sep=0.2em, s sep=0.4em, anchor=center,grow=north,}
[,circle,draw,fill  [$2$,circle,draw, ] [$3$,circle,draw [$4$,circle,draw] ]  [$1$,circle,draw]] 
\end{forest}, \hspace{0.5cm} 
\begin{forest}
for tree={l sep=0.2em, s sep=0.4em, anchor=center,grow=north}
[,circle,draw,fill [$1$,circle,draw] [$2$,circle,draw ] [$3$,circle,draw [$4$,circle,draw] ]]  
\end{forest},  
\end{center}

\noindent represent the same rooted tree. Then, forgetting the labels, the series 

\begin{eqnarray}
Z_n \left( l_1, \ldots, l_n  \right) = \sum_{k=1}^n \left[ H^{\bullet}_{0 \xrightarrow[]{n} 0} \left( \left( [1]^{j_1}, [2]^{j_2}, \ldots \right),  \left( [1]^{j_1}, [2]^{j_2}, \ldots \right) \right) \right] B_+ \left(l_1^{j_1} l_2^{j_2} \cdots l_n^{j_n} \right)
\end{eqnarray}

\noindent is a generating function that enumerates the $S_n$ permutations associated to the isomorphism classes of labelled rooted trees. To illustrate this, we consider below, the cases for $n \in [2,4]$.  

For $n=2$, we consider the permutations $\left\{ (1)(2),(12)  \right\}$ of labelled vertices in the subtrees to give

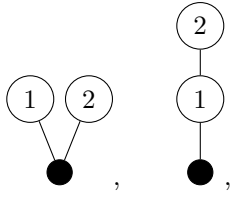
\begin{center}
\begin{forest}
for tree={l sep=0.2em, s sep=0.4em, anchor=center,grow=north,}
[,circle,draw,fill  [$2$,circle,draw, ] [$1$,circle,draw]] 
\end{forest}, \hspace{0.5cm} 
\begin{forest}
for tree={l sep=0.2em, s sep=0.4em, anchor=center,grow=north}
[,circle,draw,fill [$1$,circle,draw, [$2$,circle,draw]] ] 
\end{forest},    
\end{center}

\noindent and we write $Z_2$ as

\begin{eqnarray}
Z_2 \left( 
\begin{forest}
for tree={circle,draw,fill,minimum width=2pt,inner sep=0pt,parent anchor=center,child anchor=center,s sep+=0pt,l sep-=5pt,grow=north,}
[, for tree={l=0}  ] 
\end{forest}, \hspace{0.25cm}
\begin{forest}
for tree={circle,draw,fill,minimum width=2pt,inner sep=0pt,parent anchor=center,child anchor=center,s sep+=0pt,l sep-=5pt,grow=north,}
[, for tree={l=0} []] 
\end{forest}
\right) = \hspace{0.25cm} \frac{1}{2!} \left(
\begin{forest}
for tree={circle,draw,fill,minimum width=2pt,inner sep=0pt,parent anchor=center,child anchor=center,s sep+=0pt,l sep-=5pt,grow=north,}
[, for tree={l=0} [][] ] 
\end{forest} \hspace{0.25cm}+ \hspace{0.25cm}
\begin{forest}
for tree={circle,draw,fill,minimum width=2pt,inner sep=0pt,parent anchor=center,child anchor=center,s sep+=0pt,l sep-=5pt,grow=north,}
[, for tree={l=0} [[]]] 
\end{forest} \right) \hspace{0.25cm}.
\end{eqnarray}

For $n=3$, we consider the permutations $\left\{ (1)(2)(3),(12)(3),(13)(2),(23)(1),(123),(132)  \right\}$ of labelled vertices in the subtrees to give 

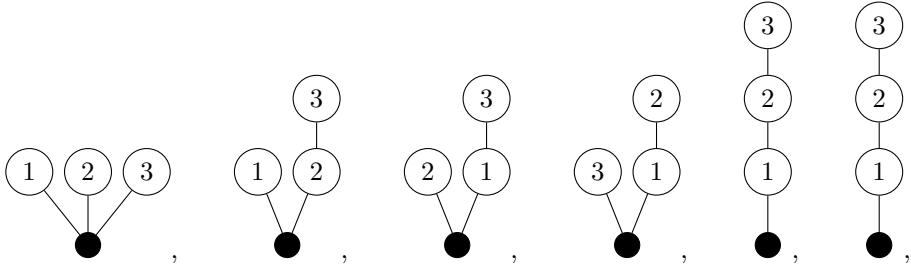
\begin{center}
\begin{forest}
for tree={l sep=0.2em, s sep=0.4em, anchor=center,grow=north,}
[,circle,draw,fill  [$3$,circle,draw, ] [$2$,circle,draw, ] [$1$,circle,draw]] 
\end{forest}, \hspace{0.5cm} 
\begin{forest}
for tree={l sep=0.2em, s sep=0.4em, anchor=center,grow=north}
[,circle,draw,fill  [$2$,circle,draw, [$3$,circle,draw]] [$1$,circle,draw,] ] 
\end{forest}, \hspace{0.5cm}    
\begin{forest}
for tree={l sep=0.2em, s sep=0.4em, anchor=center,grow=north}
[,circle,draw,fill  [$1$,circle,draw, [$3$,circle,draw]] [$2$,circle,draw,]] 
\end{forest}, \hspace{0.5cm}
\begin{forest}
for tree={l sep=0.2em, s sep=0.4em, anchor=center,grow=north}
[,circle,draw,fill  [$1$,circle,draw, [$2$,circle,draw]] [$3$,circle,draw,]] 
\end{forest}, \hspace{0.5cm}
\begin{forest}
for tree={l sep=0.2em, s sep=0.4em, anchor=center,grow=north}
[,circle,draw,fill  [$1$,circle,draw, [$2$,circle,draw [$3$,circle,draw,]]] ] 
\end{forest}, \hspace{0.5cm}
\begin{forest}
for tree={l sep=0.2em, s sep=0.4em, anchor=center,grow=north}
[,circle,draw,fill  [$1$,circle,draw, [$2$,circle,draw [$3$,circle,draw,]]] ] 
\end{forest},
\end{center}

\noindent and we write $Z_3$ as

\begin{eqnarray}
Z_3 \left( 
\begin{forest}
for tree={circle,draw,fill,minimum width=2pt,inner sep=0pt,parent anchor=center,child anchor=center,s sep+=0pt,l sep-=5pt,grow=north,}
[, for tree={l=0}  ] 
\end{forest}, \hspace{0.25cm}
\begin{forest}
for tree={circle,draw,fill,minimum width=2pt,inner sep=0pt,parent anchor=center,child anchor=center,s sep+=0pt,l sep-=5pt,grow=north,}
[, for tree={l=0} []] 
\end{forest}, \hspace{0.25cm}
\begin{forest}
for tree={circle,draw,fill,minimum width=2pt,inner sep=0pt,parent anchor=center,child anchor=center,s sep+=0pt,l sep-=5pt,grow=north,}
[, for tree={l=0} [[]]] 
\end{forest}
\right) = \hspace{0.25cm} \frac{1}{3!} \left(
\begin{forest}
for tree={circle,draw,fill,minimum width=2pt,inner sep=0pt,parent anchor=center,child anchor=center,s sep+=0pt,l sep-=5pt,grow=north,}
[, for tree={l=0} [][][]] 
\end{forest} \hspace{0.25cm} + \hspace{0.25cm} 3 \hspace{0.25cm}
\begin{forest}
for tree={circle,draw,fill,minimum width=2pt,inner sep=0pt,parent anchor=center,child anchor=center,s sep+=0pt,l sep-=5pt,grow=north,}
[, for tree={l=0} [[]] []] 
\end{forest} \hspace{0.25cm} + \hspace{0.25cm} 2 \hspace{0.25cm}
\begin{forest}
for tree={circle,draw,fill,minimum width=2pt,inner sep=0pt,parent anchor=center,child anchor=center,s sep+=0pt,l sep-=5pt,grow=north,}
[, for tree={l=0} [[[]]] ] 
\end{forest} \right) \hspace{0.25cm}.
\end{eqnarray}

For $n=4$, the subtrees' labelled vertices associated to the permutation elements of $S_4$ are

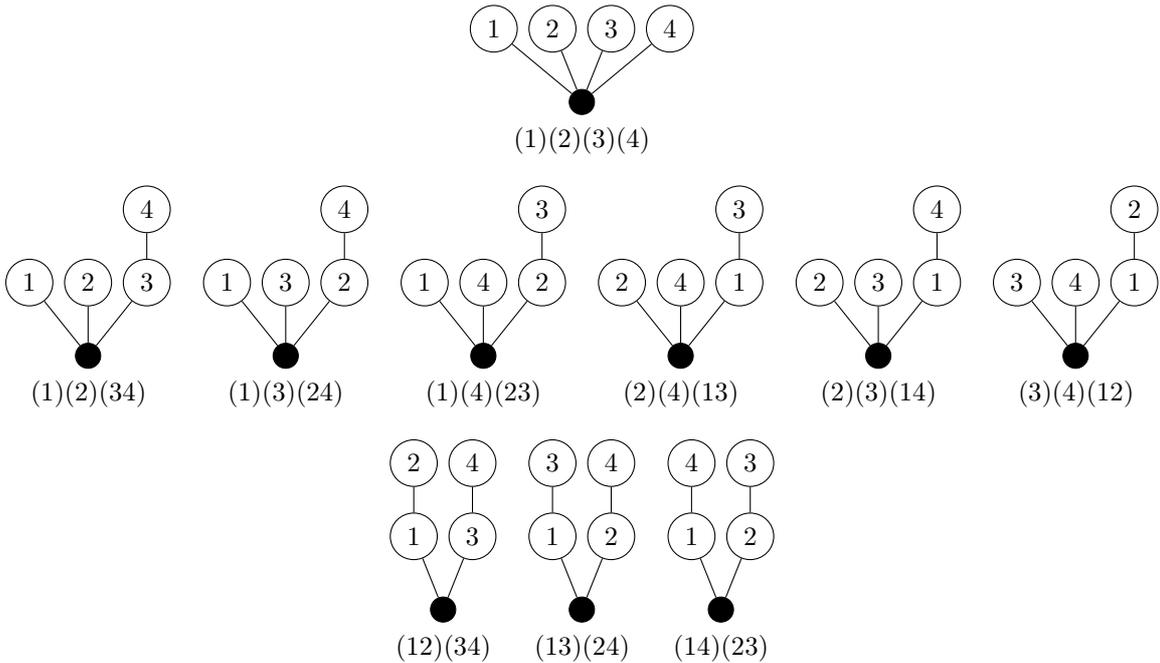
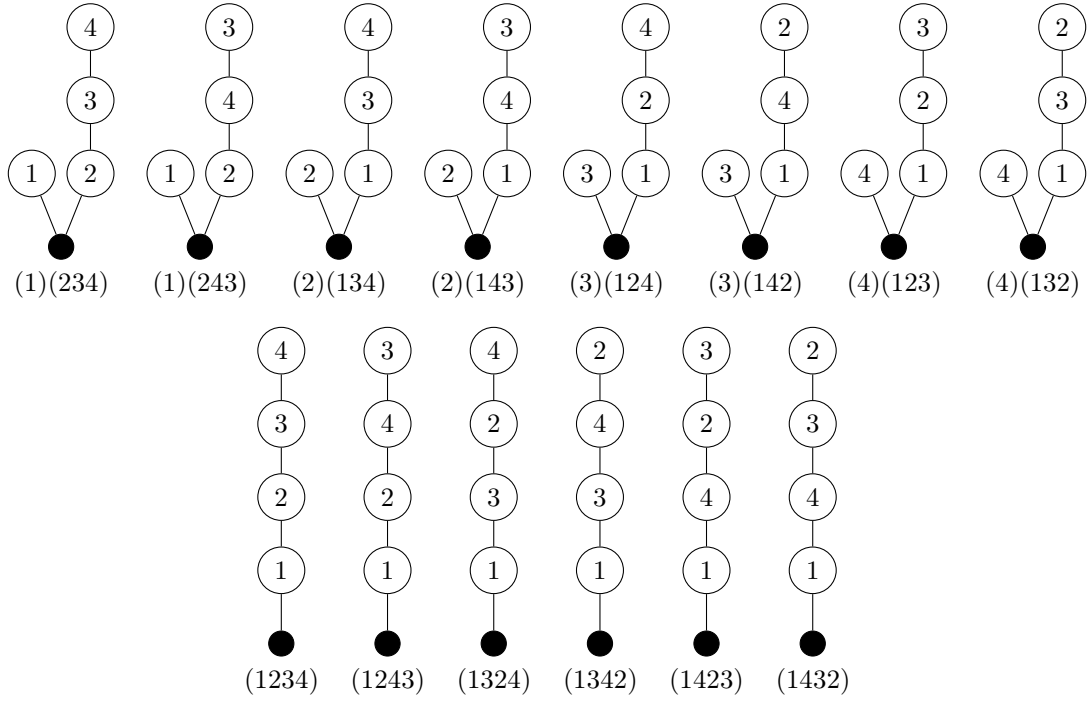
\begin{center}
\begin{tabular}{c}
 \begin{forest}
for tree={l sep=0.2em, s sep=0.4em, anchor=center,grow=north,}
[,circle,draw,fill [$4$,circle,draw, ] [$3$,circle,draw, ] [$2$,circle,draw, ] [$1$,circle,draw]] 
\end{forest} \\ 
 (1)(2)(3)(4)   
\end{tabular}
\end{center}

\begin{center}
\begin{tabular}{cccccc}
\begin{forest}
for tree={l sep=0.2em, s sep=0.4em, anchor=center,grow=north}
[,circle,draw,fill  [$3$,circle,draw, [$4$,circle,draw]] [$2$,circle,draw,] [$1$,circle,draw,] ] 
\end{forest} & 
\begin{forest}
for tree={l sep=0.2em, s sep=0.4em, anchor=center,grow=north}
[,circle,draw,fill  [$2$,circle,draw, [$4$,circle,draw]] [$3$,circle,draw,] [$1$,circle,draw,] ] 
\end{forest}&
\begin{forest}
for tree={l sep=0.2em, s sep=0.4em, anchor=center,grow=north}
[,circle,draw,fill  [$2$,circle,draw, [$3$,circle,draw]] [$4$,circle,draw,] [$1$,circle,draw,] ] 
\end{forest}&
\begin{forest}
for tree={l sep=0.2em, s sep=0.4em, anchor=center,grow=north}
[,circle,draw,fill  [$1$,circle,draw, [$3$,circle,draw]] [$4$,circle,draw,] [$2$,circle,draw,] ] 
\end{forest}&
\begin{forest}
for tree={l sep=0.2em, s sep=0.4em, anchor=center,grow=north}
[,circle,draw,fill  [$1$,circle,draw, [$4$,circle,draw]] [$3$,circle,draw,] [$2$,circle,draw,] ] 
\end{forest}&
\begin{forest}
for tree={l sep=0.2em, s sep=0.4em, anchor=center,grow=north}
[,circle,draw,fill  [$1$,circle,draw, [$2$,circle,draw]] [$4$,circle,draw,] [$3$,circle,draw,] ] 
\end{forest} \\ 
 (1)(2)(34) & (1)(3)(24) & (1)(4)(23) & (2)(4)(13) & (2)(3)(14) & (3)(4)(12)  
\end{tabular}
\end{center}

\begin{center}
\begin{tabular}{ccc}
\begin{forest}
for tree={l sep=0.2em, s sep=0.4em, anchor=center,grow=north}
[,circle,draw,fill  [$3$,circle,draw, [$4$,circle,draw]] [$1$,circle,draw, [$2$,circle,draw,]] ] 
\end{forest} & 
\begin{forest}
for tree={l sep=0.2em, s sep=0.4em, anchor=center,grow=north}
[,circle,draw,fill  [$2$,circle,draw, [$4$,circle,draw]] [$1$,circle,draw, [$3$,circle,draw,]] ] 
\end{forest}& 
\begin{forest}
for tree={l sep=0.2em, s sep=0.4em, anchor=center,grow=north}
[,circle,draw,fill  [$2$,circle,draw, [$3$,circle,draw]] [$1$,circle,draw, [$4$,circle,draw,]] ] 
\end{forest} \\
(12)(34)  & (13)(24)& (14)(23)
\end{tabular}    
\end{center}

\begin{center}
\begin{tabular}{cccccccc}
\begin{forest}
for tree={l sep=0.2em, s sep=0.4em, anchor=center,grow=north}
[,circle,draw,fill  [$2$,circle,draw, [$3$,circle,draw, [$4$,circle,draw]]] [$1$,circle,draw,] ] 
\end{forest} & 
\begin{forest}
for tree={l sep=0.2em, s sep=0.4em, anchor=center,grow=north}
[,circle,draw,fill  [$2$,circle,draw, [$4$,circle,draw, [$3$,circle,draw]]] [$1$,circle,draw,] ] 
\end{forest}& 
\begin{forest}
for tree={l sep=0.2em, s sep=0.4em, anchor=center,grow=north}
[,circle,draw,fill  [$1$,circle,draw, [$3$,circle,draw, [$4$,circle,draw]]] [$2$,circle,draw,] ] 
\end{forest}& 
\begin{forest}
for tree={l sep=0.2em, s sep=0.4em, anchor=center,grow=north}
[,circle,draw,fill  [$1$,circle,draw, [$4$,circle,draw, [$3$,circle,draw]]] [$2$,circle,draw,] ] 
\end{forest}& 
\begin{forest}
for tree={l sep=0.2em, s sep=0.4em, anchor=center,grow=north}
[,circle,draw,fill  [$1$,circle,draw, [$2$,circle,draw, [$4$,circle,draw]]] [$3$,circle,draw,] ] 
\end{forest}& 
\begin{forest}
for tree={l sep=0.2em, s sep=0.4em, anchor=center,grow=north}
[,circle,draw,fill  [$1$,circle,draw, [$4$,circle,draw, [$2$,circle,draw]]] [$3$,circle,draw,] ] 
\end{forest}& 
\begin{forest}
for tree={l sep=0.2em, s sep=0.4em, anchor=center,grow=north}
[,circle,draw,fill  [$1$,circle,draw, [$2$,circle,draw, [$3$,circle,draw]]] [$4$,circle,draw,] ] 
\end{forest}& 
\begin{forest}
for tree={l sep=0.2em, s sep=0.4em, anchor=center,grow=north}
[,circle,draw,fill  [$1$,circle,draw, [$3$,circle,draw, [$2$,circle,draw]]] [$4$,circle,draw,] ] 
\end{forest}\\
 (1)(234)    & (1)(243)& (2)(134)& (2)(143)& (3)(124)& (3)(142)& (4)(123)& (4)(132)
\end{tabular}    
\end{center}

\begin{center}
\begin{tabular}{cccccc}
\begin{forest}
for tree={l sep=0.2em, s sep=0.4em, anchor=center,grow=north}
[,circle,draw,fill  [$1$,circle,draw, [$2$,circle,draw [$3$,circle,draw,[$4$,circle,draw,]]]] ] 
\end{forest}& 
\begin{forest}
for tree={l sep=0.2em, s sep=0.4em, anchor=center,grow=north}
[,circle,draw,fill  [$1$,circle,draw, [$2$,circle,draw [$4$,circle,draw,[$3$,circle,draw,]]]] ] 
\end{forest}& 
\begin{forest}
for tree={l sep=0.2em, s sep=0.4em, anchor=center,grow=north}
[,circle,draw,fill  [$1$,circle,draw, [$3$,circle,draw [$2$,circle,draw,[$4$,circle,draw,]]]] ] 
\end{forest}& 
\begin{forest}
for tree={l sep=0.2em, s sep=0.4em, anchor=center,grow=north}
[,circle,draw,fill  [$1$,circle,draw, [$3$,circle,draw [$4$,circle,draw,[$2$,circle,draw,]]]] ] 
\end{forest}& 
\begin{forest}
for tree={l sep=0.2em, s sep=0.4em, anchor=center,grow=north}
[,circle,draw,fill  [$1$,circle,draw, [$4$,circle,draw [$2$,circle,draw,[$3$,circle,draw,]]]] ] 
\end{forest}& 
\begin{forest}
for tree={l sep=0.2em, s sep=0.4em, anchor=center,grow=north}
[,circle,draw,fill  [$1$,circle,draw, [$4$,circle,draw [$3$,circle,draw,[$2$,circle,draw,]]]] ] 
\end{forest}\\
(1234) & (1243)& (1324)& (1342)& (1423)& (1432)
\end{tabular}    
\end{center}

\noindent and we write $Z_4$ as

\begin{eqnarray}
Z_4 \left( 
\begin{forest}
for tree={circle,draw,fill,minimum width=2pt,inner sep=0pt,parent anchor=center,child anchor=center,s sep+=0pt,l sep-=5pt,grow=north,}
[, for tree={l=0}  ] 
\end{forest}, \hspace{0.25cm}
\begin{forest}
for tree={circle,draw,fill,minimum width=2pt,inner sep=0pt,parent anchor=center,child anchor=center,s sep+=0pt,l sep-=5pt,grow=north,}
[, for tree={l=0} [] ] 
\end{forest}, \hspace{0.25cm}
\begin{forest}
for tree={circle,draw,fill,minimum width=2pt,inner sep=0pt,parent anchor=center,child anchor=center,s sep+=0pt,l sep-=5pt,grow=north,}
[, for tree={l=0} [[]]] 
\end{forest}, \hspace{0.25cm}
\begin{forest}
for tree={circle,draw,fill,minimum width=2pt,inner sep=0pt,parent anchor=center,child anchor=center,s sep+=0pt,l sep-=5pt,grow=north,}
[, for tree={l=0} [[[]]]] 
\end{forest}
\right) = \hspace{0.25cm} \frac{1}{4!} \left(
\begin{forest}
for tree={circle,draw,fill,minimum width=2pt,inner sep=0pt,parent anchor=center,child anchor=center,s sep+=0pt,l sep-=5pt,grow=north,}
[, for tree={l=0} [][][][]] 
\end{forest} \hspace{0.25cm} + \hspace{0.25cm} 6 \hspace{0.25cm}
\begin{forest}
for tree={circle,draw,fill,minimum width=2pt,inner sep=0pt,parent anchor=center,child anchor=center,s sep+=0pt,l sep-=5pt,grow=north,}
[, for tree={l=0} [[]] [] []] 
\end{forest} \hspace{0.25cm} + \hspace{0.25cm} 3 \hspace{0.25cm}
\begin{forest}
for tree={circle,draw,fill,minimum width=2pt,inner sep=0pt,parent anchor=center,child anchor=center,s sep+=0pt,l sep-=5pt,grow=north,}
[, for tree={l=0} [[]] [[]]] 
\end{forest} \hspace{0.25cm} + \hspace{0.25cm} 8 \hspace{0.25cm}
\begin{forest}
for tree={circle,draw,fill,minimum width=2pt,inner sep=0pt,parent anchor=center,child anchor=center,s sep+=0pt,l sep-=5pt,grow=north,}
[, for tree={l=0} [[[]]] []] 
\end{forest} \hspace{0.25cm} + \hspace{0.25cm} 6 \hspace{0.25cm}
\begin{forest}
for tree={circle,draw,fill,minimum width=2pt,inner sep=0pt,parent anchor=center,child anchor=center,s sep+=0pt,l sep-=5pt,grow=north,}
[, for tree={l=0} [[[[]]]] ] 
\end{forest} \right) \hspace{0.25cm}.
\end{eqnarray} 

\noindent This can easily be verified to be consistent at higher order in $n$. 

The present work nicely fits with recently obtained results \cite{Mvondo-She:2021joh} where, in order to specify the genus zero Hurwitz numbers that feature in the log partition function, we considered the map $f: \left( \mathbb{C}^2 \right)^n \mapsto \left( \mathbb{C}^2 \right)^n/S_n$, and related these Hurwitz numbers to combinatorial properties of the symmetric group $S_n$ by describing an appropriate Hurwitz cover with monodromies specified by $S_n$-permutations of the $n$-sheets of Riemann surfaces. The Hurwitz numbers were therefore equivalent to the monodromies weighted by the inverse of the number of automorphisms of the Hurwitz cover, i.e $n!$. We can see from the present work that if we now consider isomorphism classes of rooted trees, according to a specific choice of permutations in the subtrees forming the rooted trees, the same Hurwitz numbers appear now as the sum of these isomorphism classes of rooted trees weighted by the inverse of the number of automorphisms of the Hurwitz cover. This highlights the fact that via Hurwitz theory, ramified coverings naturally provide an interesting correspondence between trees and (genus zero) Riemann surfaces. At level $n=2$ for instance, this correspondence is pictured in the figures below. 

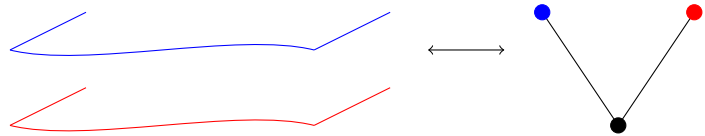
\begin{figure}[H]
\begin{center}
\begin{tikzpicture}
\draw[blue] (-7,0.5) .. controls (-6,0.25) and (-4,0.75).. (-3,0.5);
\draw[-,blue]  (-7,0.5) -- (-6,1);
\draw[-,blue]  (-3,0.5) -- (-2,1);
\draw[red] (-7,-0.5) .. controls (-6,-0.75) and (-4,-0.25).. (-3,-0.5);
\draw[-,red]  (-7,-0.5) -- (-6,0);
\draw[-,red]  (-3,-0.5) -- (-2,0);
\draw[<->] (-1.5,0.5) -- (-0.5,0.5);
\draw[-] (0,1) -- (1,-0.5);
\draw[-] (1,-0.5) -- (2,1);
\filldraw[blue] (0,1) circle (0.1);
\filldraw[black] (1,-0.5) circle (0.1);
\filldraw[red] (2,1) circle (0.1);
\end{tikzpicture}    
\end{center}
\caption{Riemann surfaces and rooted tree associated to permutation $(1)(2) \in S_2$}
\label{fig4}
\end{figure}

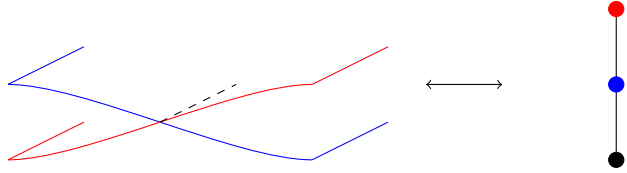
\begin{figure}[H]
\begin{center}
\begin{tikzpicture}
\draw[blue] (-7,0.5) .. controls (-6,0.5) and (-4,-0.5).. (-3,-0.5);
\draw[-,blue]  (-7,0.5) -- (-6,1);
\draw[-,blue]  (-3,-0.5) -- (-2,0);
\draw[red] (-7,-0.5) .. controls (-6,-0.5) and (-4,0.5).. (-3,0.5);
\draw[-,red]  (-7,-0.5) -- (-6,0);
\draw[-,red]  (-3,0.5) -- (-2,1);
\draw[-,dashed,black]  (-5,0) -- (-4,0.5);
\draw[<->] (-1.5,0.5) -- (-0.5,0.5);
\draw[-] (1,-0.5) -- (1,0.5);
\draw[-] (1,0.5) -- (1,1.5);
\filldraw[black] (1,-0.5) circle (0.1);
\filldraw[blue] (1,0.5) circle (0.1);
\filldraw[red] (1,1.5) circle (0.1);
\end{tikzpicture}    
\end{center}
\caption{Riemann surfaces and rooted tree associated to permutation $(12) \in S_2$}
\label{fig5}
\end{figure}

\noindent In Fig. (\ref{fig4}), the representations of covering maps are associated to the identity permutation expressed in cycle notation as (1)(2), and in Fig. (\ref{fig5}), the branched covering representations are associated with the nontrivial (12) permutation.

\section{Trees, traveling waves and disorder}
Beyond the combinatorial aspect of the work exposed above, an important objective is also to bring a connection between the introduction of disorder in AdS$_3$ as observed in \cite{Grumiller:2008qz} and the theory of disorder on trees. 

From the previous sections, it appears that the partition function $Z_{log}$ describes a Fock space geometry living on rooted trees. This result, coupled to the fact that $Z_{log}$ is also a $\tau$-function of the KP integrable hierarchy of nonlinear partial differential equations is of interest in analogy with works that have appeared in the statistical physics literature concerning disordered models defined on trees, and their relations to traveling waves. A specific case is given by the directed polymer in a random medium, whose discrete version is formulated with the lattice taken to be the Cayley tree \cite{huse1985pinning,imbrie1988diffusion,bolthausen1989note,cook1989polymers,cook1990lyapunov,derrida1990directed,cook1991finite}. Such a system, away (yet not far) from thermal equilibrium exhibit a time-scale hierarchical structure with quenched randomness at the microscopic level, and its study can be reduced to the classical statistical mechanics problem of a one-dimensional string-like object, the directed polymer on the Cayley tree (DPCT). 

In this particular example, it was discovered that traveling waves appear in disordered models on trees \cite{derrida1988polymers}, where the Cayley tree is closely connected to traveling wave solutions of a certain nonlinear partial differential equation called the Kolmogorov-Petrovsky-Piscounov (KPP) equation (also called the Fisher equation) \cite{KPP1937}.   

Our work on log gravity has been driven by the desire to have a better understanding of the combinatorics of the multi-particle excitations of the logarithmic partner, and by the resolution to unveil hidden phenomena in the theory, encoded by the partition function. With respect to that, we also note that representation of trees in the Fock space of multiparticle states has also appeared in the literature \cite{altshuler1997quasiparticle}. The analogies mentioned in this section highlight nontrivial correspondences between log gravity and aspects of statistical physics related to disorder systems, which deserve further investigation.


\section{Summary and outlook}
In this work, we gave a tree-level description of the logarithmic contribution of the partition function of topologically massive gravity at the critical point, showing that $Z_{log}$ can be expressed as a sum over rooted trees indexed by Hurwitz numbers. It was also shown how the cycle decomposition of permutations counted by the Hurwitz numbers appears on the rooted trees, bringing in the context of quantum gravity yet an illustration of the relationship between rooted trees and Hurwitz numbers already discussed in various places in the mathematical literature \cite{goulden2009short,chen2008localization,lando2010hurwitz,goulden2011moduli,dubrovin2017classical,chen2020masur}.

Our work also illustrates the relation between the algebraic structures introduced by Connes and Kreimer, the ones introduced by Connes and Moscovici and integrable hierarchies. Indeed, on one hand the Fa\`a di Bruno Hopf algebra is of the same type as the Hopf algebra of Feynman graphs. On the other hand, the Fa\`a di Bruno Hopf algebra is the maximal commutative Hopf subalgebra of the (noncommutative geometry) Hopf algebra used by Connes and Moscovici to study dffeomorphisms in a noncommutative geometry setting \cite{Figueroa:2005jr}. The reason for the link between these fields is that they use the (algebra of) composition of functions \cite{brouder2004trees}, which is precisely what the log partition function computes. Just as the Conne-Kreimer Hopf algebra of rooted trees and the Connes-Moscovici Hopf algebra of differential operators (in the one-dimensional case), the Fa\`a di Bruno Hopf algebra is not cocommutative Hopf algebra, because the coproduct acts on noncommutative spaces. This hints towards the presence of a noncommutative space of solitons in the theory.

The partition function of the log sector of TMG at the critical point is at the confluence of many theories, among which nonunitary gravity, the plethystic programme, integrable hierarchies of soliton equations and $\tau$-functions, Hurwitz theory and branched covering of Riemann surfaces and the Connes-Kreimer Hopf algebra of rooted trees and Feynman diagrams. The links between on one hand integrable hierarchies, Schur polynomials, Hurwitz numbers and matrix models  \cite{Chekhov:2005kd,deMelloKoch:2010hav,Alexandrov:2010th,Orlov:2017nub,Orlov:2018mbz,Ambjorn:2014jwa,Natanzon:2014mda,Ambjorn:2014tsa}, and on the other hand the work of Connes and Kreimer within the formalism of \cite{Gerasimov:2000pr} well suited for applications to matrix models \cite{Itoyama:2017emp,Itoyama:2017xid}, naturally brings us to question whether the fact that all the aforementioned objects that appear in the log partition function can lead to a matrix model interpretation of the counting problem in the log sector. We hope to shed some light on this matter in the future.

A more physical motivation for this work is given by the analogy between our results and the problem of the directed polymer on a tree with disorder, a notable type of disordered system which can be reduced to the study nonlinear partial differential equations that admit travelling wave solutions. The analogy comes from the fact that a hierarchical tree-like structure captured by $Z_{log}$ encodes the geometry of the Fock space of multiparticle states, bringing a relationship between trees in a disordered landscape and traveling wave solutions (of KP solitonic type in our case). This calls our attention to the fact that the sector of the theory counted by $Z_{log}$ appears as a random medium with partial equilibrium, and suggests that there might be a need to incorporate a discussion about nonequilibrium phenomena in log gravity. We report on this elsewhere.

\paragraph{Acknowledgements} The author would like to thank Sergei Chmutov and Dominique Manchon for helpful correspondence on unlabelled nonplanar rooted trees, as well as Robert de Mello Koch for discussion and comments concerning this work. This work is supported by the South African Research Chairs initiative of the Department of Science and Technology and the National Research Foundation. The support of the DSI-NRF Centre of Excellence in Mathematical and Statistical Sciences (CoE-MaSS) towards this research is hereby acknowledged. Opinions expressed and conclusions arrived at, are those of the author and are not necessarily to be attributed to the CoE.


\clearpage

\bibliographystyle{utphys}
\bibliography{main}
\end{document}